\documentclass[a4paper,11pt]{article}
\pdfoutput=1 

\usepackage{amsmath}                    
\usepackage[bbgreekl]{mathbbol}
\DeclareSymbolFontAlphabet{\mathbbl}{bbold}
\usepackage{amsfonts,amssymb,amsthm,mathtools}    
\DeclareSymbolFontAlphabet{\mathbbm}{bbold}
\DeclareSymbolFontAlphabet{\mathbb}{AMSb}%
\usepackage{tensor,mathrsfs}
\usepackage[utf8]{inputenc} 
\usepackage[T1]{fontenc}
\usepackage{geometry} 				
\usepackage{lmodern}
\usepackage{enumerate}
\usepackage{titlesec}
\usepackage{bm}
\usepackage[dvipsnames]{xcolor}            
\usepackage[customcolors]{hf-tikz}
\usepackage{array}
\usepackage{authblk}
\usepackage[numbers,sort&compress]{natbib}
\usepackage{scalerel}
\usepackage{tensor}
\usepackage{jheppub}
\usepackage{accents}
\usepackage{todonotes}
\usepackage{accents}

\newtheorem{theorem}{Theorem}[section]

\makeatletter
\let\c@equation\c@theorem
\makeatother
\usepackage{chngcntr}
\counterwithin{equation}{section}

\newtheorem{remark}[theorem]{Remark}

\newtheorem*{lemma*}{Lemma}                        
              
\renewcommand{\SS}{\mathbb{S}}
\newcommand{\Sol}{\mathrm{Sol}}
\newcommand{\corurl}{BrickRed}  \newcommand{\corcite}{red}
\newcommand{\corlink}{blue}    \newcommand{\corfile}{black}

\hypersetup{
	linktocpage,
	colorlinks,
	urlcolor=\corurl,	
	citecolor=\corcite,
	linkcolor=\corlink,
	filecolor=\corfile,
	pdfnewwindow,
	pdftitle={Holst gravity},
	pdfauthor={FB JMB ESV},
	pdfsubject={CPS}}
\usepackage{todonotes}
\usetikzlibrary{matrix}

\def\wwedgee{{\setbox0\hbox{\ensuremath{\mathrel{\wedge}}}\rlap{\hbox to \wd0{\hss\,\ensuremath\wedge\hss}}\box0}}

\newcommand{\wwedge}{\mathrel{\!\wwedgee\!}}
\def\ledgee{{\setbox0\hbox{\ensuremath{\mathrel{\cdot}}}\rlap{\hbox to \wd0{\hss\ensuremath\wedge\hss}}\box0}}

\usepackage{graphicx}
\makeatletter
\newcommand\rrule[3][0pt]{%
	\ifdim#2>#3\math@hrule[#1]{#2}{#3}\else\math@vrule[#1]{#2}{#3}\fi}
\newcommand\math@hrule[3][0pt]{%
	\gdef\mystery@factor{0.07}%
	\@tempdima=#3%
	\rule[#1]{0pt}{#3}
	\raisebox{.5\@tempdima+#1}{%
		\makebox[#2][l]{\kern-.5\@tempdima\@@mathrule{#2}{#3}}}%
}
\newcommand\math@vrule[3][0pt]{%
	\gdef\mystery@factor{0.0}%
	\@tempdima=#2%
	\rule[#1]{0pt}{#3}
	\raisebox{-.0\@tempdima+#1}{%
		\kern0.5\@tempdima%
		\rotatebox{90}{\kern-0.5\@tempdima\makebox[#3][l]{\@@mathrule{#3}{#2}}}%
		\kern0.5\@tempdima}%
}
\def\@@mathrule#1#2{%
	\@tempdimb=#2%
	\@tempdima=\dimexpr#1-\mystery@factor\@tempdimb
	\pdfliteral{%
		q []0 d %
		1 J 
		\strip@pt\@tempdimb\space w \strip@pt\@tempdimb\space 0 m %
		\strip@pt\@tempdima\space 0 l S Q }}
\makeatother

\graphicspath{{./font/}{./}}

\newcommand{\dd}{\mathchoice
	{\mathbbm{d}\rrule{.087ex}{1.605ex}\hspace*{0.15ex}} 
	{\mathbbm{d}\rrule{.087ex}{1.605ex}\hspace*{0.15ex}} 
	{\mathbbm{d}\rrule{.08ex}{1.125ex}\hspace*{0.15ex}}  
	{\mathbbm{d}\rrule{.06ex}{.8ex}\hspace*{0.15ex}}     
}

\newlength{\alturaL}
\newcommand{\LL}{\includegraphics[height=1.1\alturaL]{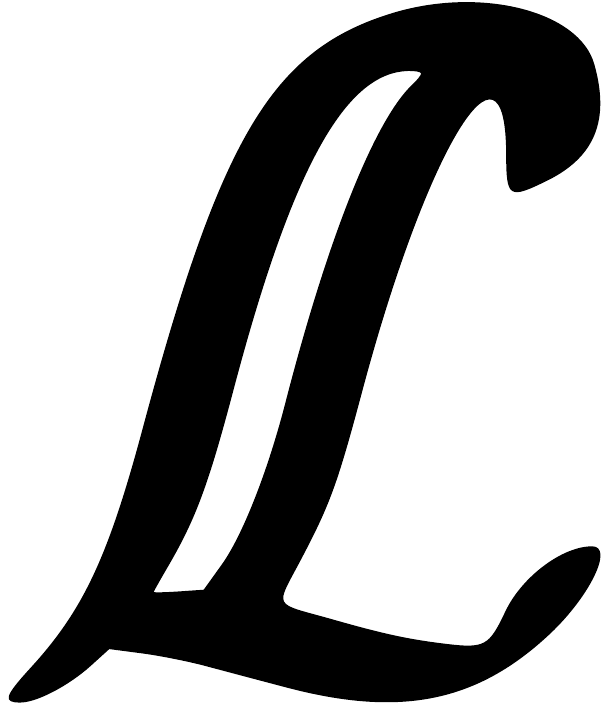}}

\newlength{\alturaO}
\newcommand{\OOmega}{\includegraphics[height=\alturaO]{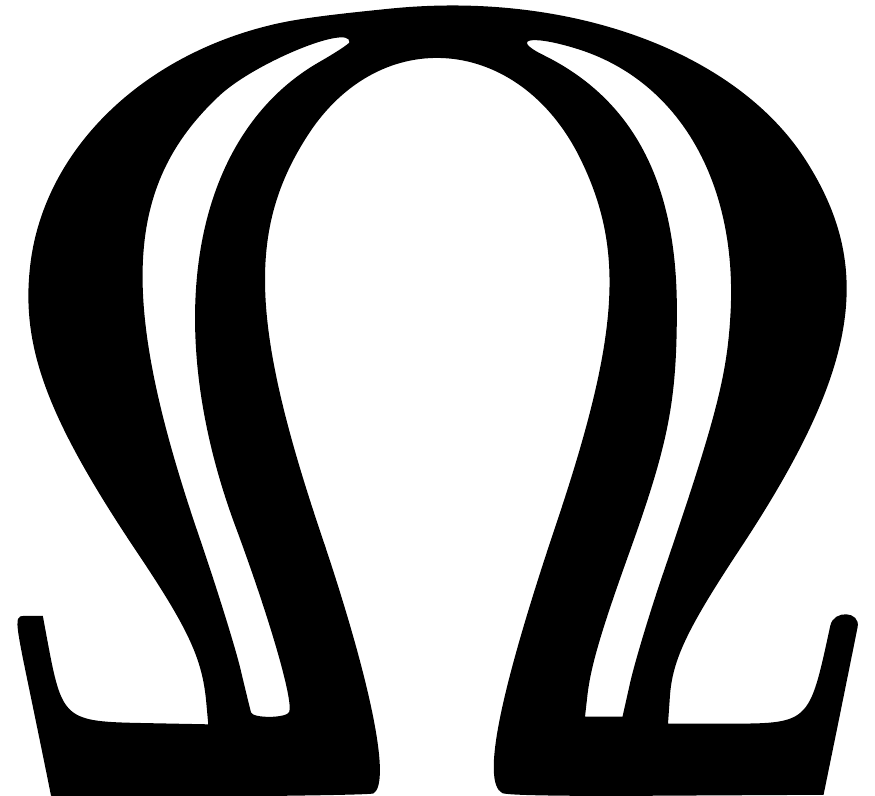}}

\newlength{\alturaI}
\newcommand{\ii}{\raisebox{-.04ex}{\includegraphics[height=1.1\alturaI]{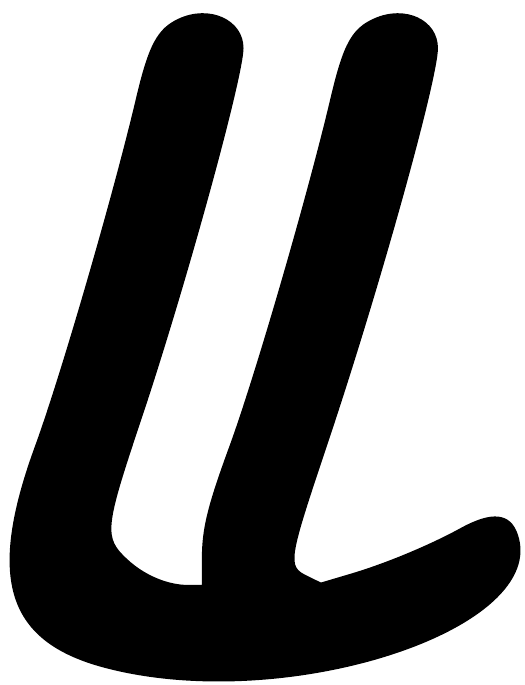}}}

\newlength{\alturaJ}
\newcommand{\jj}{\raisebox{-.37ex}{\includegraphics[height=.9\alturaJ]{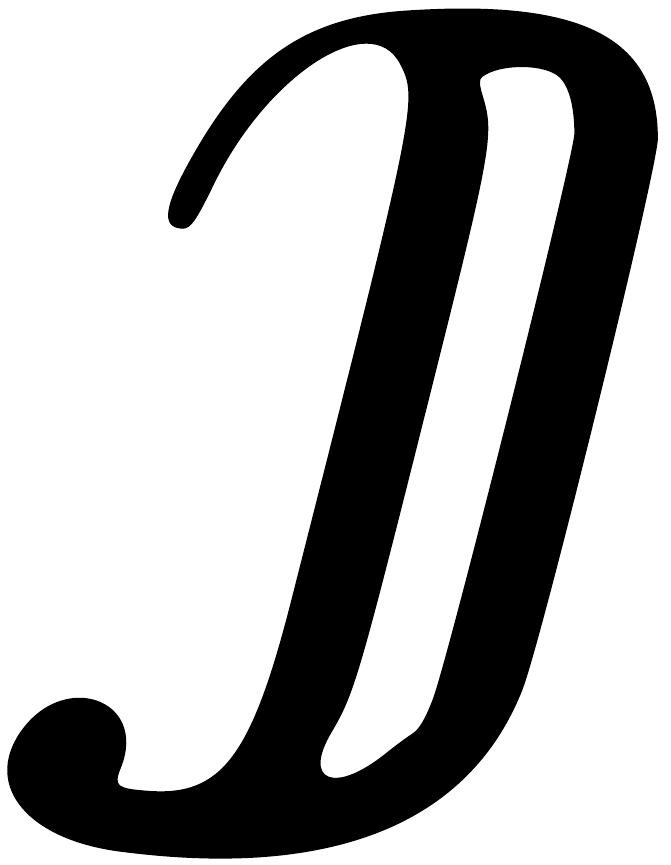}}}

\newlength{\alturaX}

\newcommand{\definicion}{\hfsetfillcolor{blue!5}\hfsetbordercolor{blue}}
\newcommand{\comentario}{\hfsetfillcolor{green!5}\hfsetbordercolor{green!50!black}}
\renewcommand{\L}{\mathcal{L}}

\newcommand{\QQ}{\mathbb{Q}}
\newcommand{\XX}{\mathbb{X}}

\newcommand{\ZZ}{\mathbb{Z}}

\newcommand{\vol}{\mathrm{vol}}

\renewcommand{\d}{\mathrm{d}}
\newcommand{\LC}[1]{\accentset{\circ}{#1}}

\usepackage{titlesec}
\usepackage{slashed}

\title{On the on-shell equivalence of general relativity and Holst theories with nonmetricity, torsion, and boundaries}

\author[c,d]{\small{J. Fernando Barbero G.}}  
\author[a,c]{\small{Juan Margalef-Bentabol}}
\author[b,c]{\small{Valle Varo}}
\author[b,c]{\small{Eduardo J.S.~Villaseñor}}

\emailAdd{fbarbero@iem.cfmac.csic.es}
\emailAdd{juanmargalef@mun.ca} 
\emailAdd{valle@cvb.es} 
\emailAdd{ejsanche@math.uc3m.es}

\affiliation[a]{Department of Mathematics and Statistics, Memorial University, St. John's, Newfoundland and Labrador A1C 5S7, Canada
	\vspace*{1ex} \mbox{}}
\affiliation[b]{Departamento de Matemáticas, Universidad Carlos III de Madrid. Avda. de la
	Universidad 30, 28911 Leganés, Spain.
	\vspace*{1ex} \mbox{}}
\affiliation[c]{Grupo de Teorías de Campos y Física Estadística. Instituto Gregorio Millán (UC3M).
	Unidad Asociada al Instituto de Estructura de la Materia, CSIC, Madrid, Spain.}
\affiliation[d]{Instituto de Estructura de la Materia, CSIC, Serrano 123, 28006, Madrid, Spain}

\abstract{We study a generalization of the Holst action where we admit nonmetricity and torsion in manifolds with timelike boundaries (both in the metric and tetrad formalism). We prove that its space of solutions is equal to the one of the Palatini action. Therefore, we conclude that the metric sector is in fact identical to GR, which is defined by the Einstein-Hilbert action. We further prove that, despite defining the same space of solutions, the Palatini and (the generalized) Holst Lagrangians are not cohomologically equal. Thus, the presymplectic structure and charges provided by the Covariant Phase Space method might differ. However, using the relative bicomplex framework, we show the covariant phase spaces of both theories are equivalent (and in fact equivalent to GR), as well as their charges, clarifying some open problems regarding dual charges and their equivalence in different formulations.}


\arxivnumber{}

\titlespacing\section{0pt}{0pt plus 4pt minus 2pt}{0pt plus 2pt minus 2pt}
\titlespacing\subsection{0pt}{8pt plus 4pt minus 2pt}{0pt plus 2pt minus 2pt}
\titlespacing\subsubsection{0pt}{8pt plus 4pt minus 2pt}{0pt plus 2pt minus 2pt}

\begin{document}

\maketitle


\section{INTRODUCTION}\label{sec:INTRODUCTION}

The Holst action \cite{Holst:1995pc} plays a very significant role in the study of the Hamiltonian formulation of general relativity (GR) in terms of real Ashtekar variables. Although proposed in a completely independent way, it has an interesting historical precedent in the work of Hojman, Mukku, and Sayed (HMS) \cite{hojman1980parity}. These authors constructed an action, that we denote $\SS^{(m)}_{\mathrm{HMS}}(g,\widetilde{\nabla})$, based on the realization that the metric Palatini action $\SS^{(m)}_{\mathrm{PT}}(g,\widetilde{\nabla})$ for vacuum gravity, which depends on a Lorentzian metric $g$ and a general metric-compatible connection $\widetilde{\nabla}$, could be modified, without changing the field equations, by adding a parity-violating term built with the help of the $g$-volume form and the Riemann tensor determined by $\widetilde{\nabla}$. \vspace*{2ex}

Given a gravitational action such as $\SS^{(m)}_{\mathrm{PT}}(g,\widetilde{\nabla})$, it is possible to build a new one in terms of a tetrad $e^I$ and a spin connection $\tensor{\widetilde{\omega}}{^I_J}$ by taking $g=\eta_{IJ}e^I\otimes e^J$ (here $\eta_{IJ}$ denotes the ``internal'' Minkowski metric) and writing the connection $\widetilde{\nabla}$ in terms of $e^I$ and $\tensor{\widetilde{\omega}}{^I_J}$ in an appropriate way. The $\tensor{\widetilde{\omega}}{_{IJ}}$ are taken to be antisymmetric i.e. $\tensor{\widetilde{\omega}}{_{IJ}}=-\tensor{\widetilde{\omega}}{_{JI}}$, a condition equivalent to the metric compatibility of $\widetilde{\nabla}$. This is the spirit of É. Cartan's approach to GR (see \cite{cattaneo2019reduced} for an interesting historical discussion). By proceeding in this way, one gets the Palatini-Cartan action $\SS^{(t)}_{\mathrm{PT}}(e,\widetilde{\omega})$, which has an important advantage over $\SS^{(m)}_{\mathrm{PT}}(g,\widetilde{\nabla})$: it allows the coupling of fermionic matter. When the previous procedure is implemented for $\SS^{(m)}_{\mathrm{HMS}}(g,\widetilde{\nabla})$, one gets the Holst action $\SS^{(t)}_{\mathrm{Holst}}(e,\widetilde{\omega})$, which is equal to $\SS^{(t)}_{\mathrm{PT}}(e,\widetilde{\omega})$ plus the so called ``dual term''. As a consequence, the coupling constant that multiplies the parity violating term of the HMS action is closely related to the Immirzi parameter $\gamma$.\vspace*{2ex}

When matter fields are not present, the equations of motion derived from the Palatini-Cartan and Holst actions are completely equivalent. This implies that the presence of the dual term does not change the Palatini space of solutions in a significant way \cite{rovelli1998immirzi,BarroseSa:2000vx,Montesinos:2001ww,samuel2001comment,Alexandrov:2000jw, Alexandrov2003,Ashtekar:2004eh, freidel2005pure, Oliveri2020, concise,diaz2021hamiltonian}. When gravity is coupled to bosonic matter fields, the coupling terms are \textit{independent of the connection} and therefore the field equations remain unchanged (see, for instance, \cite{jacobson1988fermions}). However, when coupled to fermionic matter fields, the  critical points of the Holst action are no longer equivalent to those of the Cartan-Palatini action  \cite{perez2006physical,freidel2005torsion, Mercuri:2006um, Bojowald:2007nu, date2009topological, banerjee2010some, shapiro2014quantum} due to the presence of the dual term. This may have important consequences. In particular, it is clear that the role of $\gamma$ in general relativity differs in a significant way from the one of the $\theta$ parameter in QCD.\vspace*{2ex}

The introduction of a connection as a dynamical variable adds several new geometric ingredients to the formulation of gravitational theories because, generically, connections will have non-vanishing torsion and will not be metric compatible. The interest in gravitational theories which allow both for torsion and nonmetricity dates back to the 1970s when  Hehl, Trautman, and their collaborators developed Cartan's theory of gravitation \cite{trautman1972einstein,hehl1973spin}. In their approach, the field equations do not determine the connection uniquely, and the condition $\widetilde{\nabla}_\alpha g_{\beta\gamma}=0$ has to be added.  In \cite{Sandberg:1975db}, Sandberg studied from a variational point of view the situation in which torsion is allowed and in which the metric compatibility with the connection is not assumed in general. Latter in the 1990s, Floreanini and Percacci \cite{Floreanini:1990kt}  considered a completely general  $GL(4)$-invariant Palatini formulation of GR in which the conditions of metricity and torsionlessness are both obtained as dynamical equations by adding appropriate terms to the action.\vspace*{2ex}

The study of the Holst action in the tetrad formalism with torsion (but assuming metricity) was done in \cite{Pons2012} in the absence of boundaries. In the presence of boundaries, significant work has been conducted to understand isolated horizons \cite{ashtekar2004isolated, AshtekarCorichi,Chatterjee:2008if, Engle:2009vc} and general surface terms in \cite{Corichi:2010ur, Wieland:2010ec, Pranzetti2016}. The equivalence of the Holst action field theories coupled to matter with torsion, nonmetricty and boundaries, remains an unexplored topic \cite{date2009topological}.\vspace*{2ex}

From a physical perspective, the presence of the dual term may modify the conserved charges of GR. The computation of these ``dual charges'' has been recently considered in \cite{Oliveri2020}, where the authors relied on symplectic and cohomological methods. They show that, without dressing the standard presymplectic potential in a suitable way, neither the Hamiltonian nor the Noether charges written in tetrad variables match the corresponding metric ones. As a consequence, the problem of figuring out which approach is physically relevant to elucidate if the charges in the metric and tetrad formalism are equivalent is still open.\vspace*{2ex}

In this paper we will study the Holst theory in all generality, i.e. allowing for torsion and nonmetricity---in \textit{both} metric and tetrad variables---and in manifolds with boundaries. The purpose of the present work is threefold:
\begin{enumerate}
\item Study the solution spaces of the metric-HMS and tetrad-HMS actions\footnote{The action $\SS^{(m)}_{\mathrm{HMS}}(g,\widetilde{\nabla})$ proposed by HMS \cite{hojman1980parity} was defined in terms of a metric-compatible connection with torsion. Nonetheless, we will refer to its generalization with both nonmetricity and torsion also as the HMS-action $\SS^{(m)}_{\mathrm{HMS}}(g,\widetilde{\nabla})$ and its tetrad counterpart as $\SS^{(t)}_{\mathrm{HMS}}(e,\widetilde{\omega})$.} and compare them with the ones obtained in other approaches. As we will see, the presence of the parity breaking terms does not change the solution spaces with respect to the ones corresponding to the original Palatini models. In \cite{CPSPT}, the metric-Palatini action was proved to give GR on the metric sector, hence the metric sector of the HMS theory is also necessarily GR. In \cite{CPSGR}, the metric and tetrad formulations were proved to be equivalent, so the same result holds for the tetrad formulation.
\item Study boundary terms for the different approaches. We derive a new boundary Lagrangian that guarantees the equivalence of the models considered. This is done in metric-connection variables and we show that, on translating the boundary term to tetrad-spin connection variables, we recover the one given in \cite{bodendorfer2013imaginary} (obtained by assuming nonmetricity).
\item Study the charges in all the cases (including ``dual charges'') and show that the cohomological approach provided by the relative bicomplex framework leads to their equivalence.
\end{enumerate}

This paper will strongly rely on the relative bicomplex framework \cite{CPS}, which provides a clean, consistent, and ambiguity-free procedure to obtain the space of solutions, the presymplectic forms canonically associated with the actions, and some relevant charges within the covariant phase space (CPS) framework.\vspace*{2ex}

In the following we consider a 4-dimensional spacetime manifold $M$ diffeomorphic to $\Sigma\times\mathbb{R}$, where $\Sigma$ is a 3-dimensional manifold with boundary $\partial\Sigma$ (possibly empty).  We will refer to $\partial_LM\cong\partial\Sigma\times\mathbb{R}$ as the \textit{lateral boundary} of $M$ and restrict ourselves to the open set of metrics making $\partial_LM$ time-like. Greek letters will denote abstract indices for tensorial objects in $M$ and barred Greek indices will be used for tensors on $\partial_LM$ (quite often the object itself will also carry a bar). The inclusion map will be denoted as $\jmath:\partial_L M\hookrightarrow M$ and its tangent map as $\jmath^\alpha_{\overline{\alpha}}$.

\section{THE GEOMETRIC ARENA}

Given a connection $\widetilde{\nabla}$, we define its torsion, Riemann, and Ricci tensors as
\begin{align*}
    &\tensor{\widetilde{\mathrm{T}}\mathrm{or}}{^\alpha_\mu_\nu}(\d\phi)_\alpha :=-[\widetilde{\nabla}_{\!\mu},\widetilde{\nabla}_{\!\nu}]\phi\,,\\
    &\tensor{\widetilde{\mathrm{R}}\mathrm{iem}}{^\alpha_\beta_\mu_\nu}Z^\beta :=([\widetilde{\nabla}_{\!\mu},\widetilde{\nabla}_{\!\nu}]+\tensor{\widetilde{\mathrm{T}}\mathrm{or}}{^\beta_\mu_\nu}\widetilde{\nabla}_{\!\beta})Z^\alpha\,,\\
    &\widetilde{\mathrm{R}}\mathrm{ic}_{\beta\nu}:=\tensor{\widetilde{\mathrm{R}}\mathrm{iem}}{^\mu_\beta_\mu_\nu}\,. 
\end{align*}
If we endow $M$ with a connection $\widetilde{\nabla}$ and a metric $g$, we can define the nonmetricity tensor, the $(g,\!\widetilde{\nabla})$-scalar-curvature,  the  $(g,\!\widetilde{\nabla})$-extrinsic-curvature of $\partial_LM$, and its trace
\begin{align*}
    &\widetilde{M}_{\alpha\beta\gamma}:=\widetilde{\nabla}_{\!\alpha}g_{\beta\gamma}\,,&&\widetilde{R}:=g^{\alpha\beta}\widetilde{\mathrm{R}}\mathrm{ic}_{\alpha\beta}\,,\\
    &\widetilde{K}_{\overline{\alpha}\overline{\beta}}:=\frac{1}{2}\jmath^\alpha_{\overline{\alpha}}\jmath^\beta_{\overline{\beta}}\Big(\widetilde{\nabla}_{\!\alpha}\nu_\beta+g_{\alpha\gamma}\widetilde{\nabla}_{\!\beta}\nu^\gamma\Big),
    &&\widetilde{K}:=\overline{g}^{\overline{\alpha}\overline{\beta}}\widetilde{K}_{\overline{\alpha}\overline{\beta}}\,,
\end{align*}
where  $\overline{g}:=\jmath^*g$ is the metric induced on $\partial_LM$, $\nu^\alpha$ the outward unit vector normal to $\partial_LM$, and $\nu_\beta:=g_{\beta\gamma}\nu^\gamma$. Notice that $\widetilde{R}$ and $\widetilde{K}_{\overline{\alpha}\overline{\beta}}$ are generalizations of the $g$-scalar $\LC{R}$ and $g$-extrinsic curvature $\LC{K}$ defined by the $g$-Levi-Civita connection $\LC{\nabla}$.\vspace*{2ex}

Given two connections $\nabla$ and $\widetilde{\nabla}$, their difference is a $(2,1)$-tensor $Q\equiv\widetilde{\nabla}-\nabla$. For a $(1,1)$-tensor $\tensor{S}{^\beta_\gamma}$ we have
\[(\widetilde{\nabla}_{\!\alpha}-\nabla_{\!\alpha})\tensor{S}{^\beta_\gamma}=\tensor{Q}{^\beta_\alpha_\mu}\tensor{S}{^\mu_\gamma}-\tensor{Q}{^\mu_\alpha_\gamma}\tensor{S}{^\beta_\mu}\,,\]
and analogously for higher order objects. It is easy to check that the following equality holds
\begin{align}\label{eq: Q=Tor+M}\begin{split}
Q_{\alpha\beta\gamma}&=\frac{1}{2}\Big(\mathrm{\widetilde{T}or}_{\alpha\beta\gamma}-\mathrm{Tor}_{\alpha\beta\gamma}-\mathrm{\widetilde{T}or}_{\beta\gamma\alpha}+\mathrm{Tor}_{\beta\gamma\alpha}
+\mathrm{\widetilde{T}or}_{\gamma\alpha\beta}-\mathrm{Tor}_{\gamma\alpha\beta}\Big)\\
&+\frac{1}{2}\Big(\widetilde{M}_{\alpha\beta\gamma}-M_{\alpha\beta\gamma}-\widetilde{M}_{\beta\gamma\alpha}+M_{\beta\gamma\alpha}
-\widetilde{M}_{\gamma\alpha\beta}+M_{\gamma\alpha\beta}\Big)\,.
\end{split}
\end{align}
From the definition of $Q$, we have the following relations between geometric objects associated with $\widetilde{\nabla}$ and $\nabla$
\begin{align}
&\tensor{\widetilde{\mathrm{R}}\mathrm{iem}}{^\alpha_\beta_\mu_\nu}=\nabla_{\!\mu}\tensor{Q}{^\alpha_\nu_\beta}+\tensor{Q}{^\alpha_\mu_\sigma}\tensor{Q}{^\sigma_\nu_\beta}-(\mu\leftrightarrow\nu)+\tensor{\mathrm{Riem}}{^\alpha_\beta_\mu_\nu}+\tensor{\mathrm{Tor}}{^\sigma_\mu_\nu}\tensor{Q}{^\alpha_\sigma_\beta}\,,\label{eq: Riem=Riem+Q}\\
&\tensor{\widetilde{\mathrm{T}}\mathrm{or}}{^\gamma_\alpha_\beta}=\tensor{\mathrm{Tor}}{^\gamma_\alpha_\beta}+\tensor{Q}{^\gamma_\alpha_\beta}-\tensor{Q}{^\gamma_\beta_\alpha}\,,\\
&\tensor{\widetilde{\mathrm{R}}\mathrm{ic}}{_\beta_\nu}=\tensor{\mathrm{Ric}}{_\beta_\nu}+\nabla_{\!\alpha}\tensor{Q}{^\alpha_\nu_\beta}+\tensor{Q}{^\alpha_\alpha_\sigma}\tensor{Q}{^\sigma_\nu_\beta}-\nabla_{\!\nu}\tensor{Q}{^\alpha_\alpha_\beta}-\tensor{Q}{^\alpha_\nu_\sigma}\tensor{Q}{^\sigma_\alpha_\beta}+\tensor{\mathrm{Tor}}{^\sigma_\alpha_\nu}\tensor{Q}{^\alpha_\sigma_\beta}\,,\\
&\widetilde{M}_{\alpha\beta\gamma}=M_{\alpha\beta\gamma}-Q_{\beta\alpha\gamma}-Q_{\gamma\alpha\beta}\,,\\
&\widetilde{R}=R+\nabla_{\!\alpha}\Big(\tensor{Q}{^\alpha^\beta_\beta}-\tensor{Q}{_\beta^\beta^\alpha}\Big)+\tensor{Q}{^\alpha_\alpha_\sigma}\tensor{Q}{^\sigma^\beta_\beta}-\tensor{Q}{^\alpha^\beta^\sigma}\tensor{Q}{_\sigma_\alpha_\beta}+\tensor{\mathrm{Tor}}{^\sigma^\alpha^\beta}\tensor{Q}{_\alpha_\sigma_\beta}\,,\label{eq: R=R+Q}\\
&\widetilde{K}_{\overline{\alpha}\overline{\beta}}=K_{\overline{\alpha}\overline{\beta}}+\frac{1}{2}\jmath^\alpha_{\overline{\alpha}}\jmath^\beta_{\overline{\beta}}\big(Q_{\alpha\beta\mu}-Q_{\mu\alpha\beta}\big)\nu^\mu\,,\\
&\widetilde{K}=K+\frac{1}{2}\big(\tensor{Q}{^\beta_\beta_\mu}-\tensor{Q}{_\mu_\beta^\beta}\big)\nu^\mu\,.\label{eq: K=K+Q}
\end{align}

By fixing a fiducial connection, usually the $g$-Levi-Civita one $\LC{\nabla}$, we can establish a bijection between connections $\widetilde{\nabla}$ and $(2,1)$-tensors $Q$. Working with the vector space of tensors is usually easier than working with the affine space of connections. This is why, in the following, we will use the variables $(g,Q)$ instead of the equivalent ones $(g,\!\widetilde{\nabla})$.\vspace*{2ex}

In order to describe the solution space and the presymplectic form, we will use the CPS algorithm \cite{CPS}, which essentially consists in introducing a pair of bulk and boundary Lagrangians, compute their variations, extract the equations of motion and symplectic potentials, and get the presymplectic form on the space of solutions. The power of this method lies in its cohomological nature, which renders it ambiguity-free: we can pick any representative Lagrangians and symplectic potentials to describe the solution spaces and compute the presymplectic form.

\section{METRIC-HMS WITHOUT BOUNDARY}\label{sec:METRICHMS}

\subsection{The action}
We consider actions of the form
\[
\mathbb{S}=\int_ML-\int_{\partial_L M}\overline{\ell}\,,
\]
defined in terms of a locally constructed bulk Lagrangian $L$ and a locally constructed boundary Lagrangian $\overline{\ell}$. In this section we assume $\partial_L M=\varnothing$, so the second integral vanishes. In the next section we will consider the case $\partial_L M\neq\varnothing$. The metric-GR, metric-Palatini, and metric-HMS actions are respectively defined by the Lagrangians
\begin{align*}
    &L^{(m)}_{\textrm{EH}}(g):=\big(\LC{R} - 2\Lambda \big)\vol\,,&&g\in\mathcal{F}^{(m)}_{\textrm{GR}}:=\{g\ |\ \jmath^*g \text{ is timelike}\}\,,\\
    &L^{(m)}_{\textrm{PT}}(g, Q):= \big(\widetilde{R} - 2\Lambda \big)\vol\,,&&(g,Q)\in\mathcal{F}^{(m)}_{\textrm{PT}}:=\mathcal{F}^{(m)}_{\textrm{GR}}\times\mathfrak{T}^2_1\,,\\
    &L^{(m)}_{\textrm{HMS}}(g, Q):=L^{(m)}_{\textrm{PT}}(g, Q)-\frac{1}{2\gamma}\vol^{\alpha\beta\mu\nu}\widetilde{\mathrm{R}}\mathrm{iem}_{\alpha\beta\mu\nu}\vol\,,&&(g,Q)\in\mathcal{F}^{(m)}_{\textrm{HMS}}:=\mathcal{F}^{(m)}_{\textrm{PT}}\,.
\end{align*}
Where, to ease the notation, we denote $\vol:=\vol_g$ the $g$-volume form. Using \eqref{eq: Riem=Riem+Q}, \eqref{eq: R=R+Q}, and the first Bianchi identity, it is possible to split the HMS Lagrangian into the standard GR term, some coupling terms and an exact form:
\begin{align}\label{eq: L=L+L+d something}
    &L^{(m)}_{\textrm{HMS}}(g,Q)=L^{(m)}_{\textrm{EH}}(g)+\hat{L}^{(m)}_{\textrm{P-CP}}(g,Q)-\frac{1}{\gamma}\hat{L}^{(m)}_{\textrm{H-CP}}(g,Q)+ \d \big(\iota_{\vec{A} - \vec{C}+\vec{q}/\gamma}\vol\big)\,,
\end{align}
where 
\begin{align}\label{eq: coupling temrs}
    \begin{array}{l}\hat{L}^{(m)}_{\textrm{P-CP}}(g,Q):=\big(C_{\lambda}A^{\lambda} - \tensor{Q}{^\alpha^\beta^\lambda}\tensor{Q}{_\lambda_\alpha_\beta}\big) \vol \,,\\[1ex]
    \hat{L}^{(m)}_{\textrm{H-CP}}(g,Q):=\vol^{\alpha\beta\mu\nu}Q_{\alpha\mu\sigma}\tensor{Q}{^\sigma_\nu_\beta} \vol\,,\end{array}\qquad
     \begin{array}{l}A^\alpha:=g^{\beta\gamma}\tensor{Q}{^\alpha_\beta_\gamma}\,,\\ B_\beta:=\tensor{Q}{^\mu_\beta_\mu}\,,\\ C_\gamma:=\tensor{Q}{^\mu_\mu_\gamma}\,,\\
     q^\mu:=\vol^{\mu\alpha\beta\nu}Q_{\alpha\beta\nu}\,.\end{array}
\end{align}

\subsection{Variations}
The variation of each Lagrangian is given by (see \cite{CPSGR,CPSPT})
\begin{align*}
   & \dd L^{(m)}_{\textrm{EH}}=(\mathfrak{E}_{\textrm{EH}}^{(m)})^{\alpha \beta} \dd g_{\alpha \beta} + \d\Theta^{(m)}_{\textrm{EH}}\,,\\
   & \dd \hat{L}^{(m)}_{\textrm{P-CP}}=(\mathfrak{E}_{\textrm{P-CP}}^{(m)})^{\alpha \beta} \dd g_{\alpha \beta} + \tensor{(\mathcal{E}_{\textrm{P-CP}}^{(m)})}{_\alpha^\beta^\sigma} \dd \tensor{Q}{^\alpha_\beta_\sigma}\,,\\
   & \dd \hat{L}^{(m)}_{\textrm{H-CP}}=(\mathfrak{E}_{\textrm{H-CP}}^{(m)})^{\alpha \beta} \dd g_{\alpha \beta} +  \tensor{(\mathcal{E}_{\textrm{H-CP}}^{(m)})}{_\alpha^\beta^\sigma} \dd \tensor{Q}{^\alpha_\beta_\sigma}\,,
\end{align*}
where
\begin{align*}
   &(\mathfrak{E}_{\textrm{EH}}^{(m)})^{\alpha \beta}:=-\left(\mathrm{\LC{R}ic}^{\alpha\beta}-\frac{1}{2}(\LC{R}-2\Lambda)g^{\alpha\beta}\right)\vol  \,,\\
   & (\mathfrak{E}_{\textrm{P-CP}}^{(m)})^{\alpha \beta}:=\frac{1}{2}\Big( \tensor{Q}{^\gamma^\alpha_\sigma}\tensor{Q}{^\sigma_\gamma^\beta} - C_{\sigma}\tensor{Q}{^\sigma^\alpha^\beta}  + \frac{1}{2}g^{\alpha \beta}(C_{\sigma}A^{\sigma} - \tensor{Q}{^\gamma^\tau_\sigma}\tensor{Q}{^\sigma_\gamma_\tau})+(\alpha\leftrightarrow\beta)\Big)\vol\,,\\
   &(\mathfrak{E}_{\textrm{H-CP}}^{(m)})^{\alpha \beta}:=\frac{1}{2}\vol^{\mu\nu\eta\xi}\Big(\delta^\beta_\xi(\tensor{Q}{_\mu^\alpha^\sigma}Q_{\sigma\eta\nu}-\tensor{Q}{_\mu_\nu^\sigma}\tensor{Q}{_\sigma_\eta^\alpha})+\delta^\alpha_\eta Q_{\mu\xi\sigma}\tensor{Q}{^\sigma^\beta_\nu}+g^{\alpha\beta}Q_{\mu\eta\sigma}\tensor{Q}{^\sigma_\xi_\nu}+(\alpha\leftrightarrow\beta)\Big)\vol\,,\\
     &\tensor{(\mathcal{E}_{\textrm{P-CP}}^{(m)})}{_\alpha^\beta^\sigma}:=  \Big( \delta^{\beta}_{\alpha}A^{\sigma} + g^{\beta\sigma}C_{\alpha} - \tensor{Q}{^\sigma_\alpha^\beta} - \tensor{Q}{^\beta^\sigma_\alpha}\Big) \vol\,,\\
     & \tensor{(\mathcal{E}_{\textrm{H-CP}}^{(m)})}{_\alpha^\beta^\sigma}:= \vol^{\xi\mu\beta\nu}\Big(g_{\xi\alpha}\tensor{Q}{^\sigma_\nu_\mu}-\delta^\sigma_\xi Q_{\nu\mu\alpha}\Big)\vol\,,\\
     &\Theta^{(m)}_{\mathrm{EH}}:=\iota_{\vec{W}}\vol\,,\quad\mathrm{with}\quad W^\alpha:=\big(g^{\alpha\mu}g^{\beta\lambda}-g^{\alpha\lambda}g^{\beta\mu}\big)\nabla_\lambda \dd g_{\beta\mu}\,.
\end{align*}
Notice that since the coupling terms have no derivatives, they do not contribute to the symplectic potentials. Gathering everything together we obtain
\[\dd L^{(m)}_{\textrm{HMS}}=\Big(\!\overbrace{\mathfrak{E}_{\textrm{EH}}^{(m)}+\mathfrak{E}_{\textrm{P-CP}}^{(m)}-\frac{1}{\gamma}\mathfrak{E}_{\textrm{H-CP}}^{(m)}\!}^{\quad\scalebox{1}{$\mathfrak{E}^{(m)}$}}\Big){\rule{0ex}{2.2ex}}^{\!\alpha \beta} \dd g_{\alpha \beta} +\Big(\!\overbrace{\mathcal{E}_{\textrm{P-CP}}^{(m)}-\frac{1}{\gamma}\mathcal{E}_{\textrm{H-CP}}^{(m)}\!}^{\quad\scalebox{1}{$\mathcal{E}^{(m)}$}}\Big)\tensor{\rule{0ex}{2.2ex}}{_{\!\alpha}^\beta^\sigma} \dd \tensor{Q}{^\alpha_\beta_\sigma}+ \d\Theta^{(m)}_{\textrm{HMS}}\]
where the following representative has been chosen as the symplectic potential (see \cite{CPSGR,CPSPT}) 
\begin{equation}\label{eq: Theta CP}
    \Theta^{(m)}_{\mathrm{HMS}}:=\Theta^{(m)}_{\mathrm{PT}}+ \frac{1}{\gamma}\dd \iota_{\vec{q}}\vol=\Theta^{(m)}_{\mathrm{EH}}+ \dd (\iota_{\vec{A}-\Vec{C}+\vec{q}/\gamma}\vol) 
\end{equation}

\subsection{Space of solutions}\label{subsection: space solution metric no boundary}
The goal of this section is to solve the algebraic equation of motion $\mathcal{E}^{(m)}=0$ for $Q$.  By plugging this solution into $\mathfrak{E}^{(m)}=0$, we will recover the Einstein equations $\mathfrak{E}_{\textrm{EH}}^{(m)}=0$. This will prove that the space of solutions of $L^{(m)}_{\textrm{HMS}}$ and $L^{(m)}_{\textrm{PT}}$ are the same.

\subsubsection*{Irreducible decomposition of the torsion}
In order to solve the first equation of motion, we will use the irreducible decomposition of the torsion and the metricity. An irreducible decomposition breaks a tensor into simpler objects with the same symmetries. Some of the terms are always built out of traces while the rest are traceless. For the torsion, which is antisymmetric in the last two indices, one obtains
\begin{equation}\label{eq: Tor=irred}
    \tensor{\widetilde{\mathrm{T}}\mathrm{or}}{^\alpha_\beta_\sigma}=\tensor{\widetilde{\tau}}{^\alpha_\beta_\sigma}+\tensor{\widetilde{\mathcal{A}}}{^\alpha_\beta_\sigma}+\tensor{\widetilde{t}}{^\alpha_\beta_\sigma}
\end{equation}
The first term is a combination of the trace $\widetilde{T}_\beta:=\tensor{\widetilde{\mathrm{T}}\mathrm{or}}{^\alpha_\beta_\alpha}$, the second term is the completely antisymmetric part of $\widetilde{\mathrm{T}}\mathrm{or}$, and the last one is the remainder components of the tensor:
\begin{equation}\label{eq: Tor decomposition}\tensor{\widetilde{\tau}}{^\alpha_\beta_\sigma}:=\frac{1}{3}(\widetilde{T}_\beta \delta^\alpha_\sigma-\widetilde{T}_\sigma \delta^\alpha_\beta)\,,\qquad\tensor{\widetilde{\mathcal{A}}}{^\alpha_\beta_\sigma}:=g^{\alpha\mu}\widetilde{\mathrm{T}}\mathrm{or}_{[\mu\beta\sigma]}\,,\qquad\tensor{\widetilde{t}}{^\alpha_\beta_\sigma}:=\tensor{\widetilde{\mathrm{T}}\mathrm{or}}{^\alpha_\beta_\sigma}-\tensor{\widetilde{\tau}}{^\alpha_\beta_\sigma}-\tensor{\widetilde{\mathcal{A}}}{^\alpha_\beta_\sigma}\,.
\end{equation}

It is easy to check that  $\widetilde{\tau},\widetilde{\mathcal{A}},\widetilde{t}$ are all antisymmetric in the last two indices. Besides, $\widetilde{\mathcal{A}},\widetilde{t}$ are traceless and $\widetilde{t}$ satisfies the cyclic identity
\begin{equation}\label{eq: t antisym=0}
\widetilde{t}_{\alpha\beta\sigma}+\widetilde{t}_{\sigma\alpha\beta}+\widetilde{t}_{\beta\sigma\alpha}=0\qquad\longrightarrow\qquad\widetilde{t}_{[\alpha\beta\sigma]}=0
\end{equation}
Finally, since the dimension of the manifold $M$ is $4$ and $\widetilde{\mathcal{A}}$ is essentially a $3$-form, we can define its dual $1$-form
\[\widetilde{d}_\mu:=4\widetilde{\mathcal{A}}^{\alpha\beta\sigma}\vol_{\alpha\beta\sigma\mu},\]
Notice that we have the relation $\widetilde{d}^\mu=-8q^\mu$ (see equation \eqref{eq: coupling temrs}). We have obtained a decomposition of any $3$-tensor antisymmetric in its two last indices given by \eqref{eq: Tor decomposition}. Conversely, we have that this decomposition completely characterizes the tensor:

\begin{remark}\mbox{}\\
The irreducible decomposition of a $3$-tensor antisymmetric in its last two indices is given by two $1$-forms $(T_\beta,d_\beta)$ and a $3$-tensor $\tensor{t}{^\alpha_\beta_\sigma}$ satisfying
\begin{equation}\label{eq: properties t}
t_{\alpha\beta\sigma}=-t_{\alpha\sigma\beta}\,,\qquad\qquad \tensor{t}{^\alpha_\alpha_\sigma}=0\,,\qquad\qquad t_{\alpha\beta\sigma}+t_{\sigma\alpha\beta}+t_{\beta\sigma\alpha}=0\,.
\end{equation}
\end{remark}
To prove that, notice that a $3$-tensor with this symmetry has $n\frac{n(n-1)}{2}$ components (equal to 24 when $n=4$) and the number of independent components of $(T_\beta,d_\beta,\tensor{t}{^\alpha_\beta_\sigma})$ is
\[n+n+\frac{1}{3}n(n^2-n)\,,\]
which is also equal to 24 when $n=4$. This proves, as intended, that $(\widetilde{T}_\beta,\widetilde{d}_\beta,\tensor{\widetilde{t}}{^\alpha_\beta_\sigma})$ is the irreducible decomposition of $\tensor{\widetilde{\mathrm{T}}\mathrm{or}}{^\alpha_\beta_\sigma}$.

\subsubsection*{Irreducible decomposition of the nonmetricity}
The nonmetricity, which is symmetric in the last two indices, has two traces $\widetilde{a}_\alpha:=\tensor{\widetilde{M}}{_\alpha_\beta^\beta}$ and $\widetilde{b}_\beta:=\tensor{\widetilde{M}}{^\alpha_\beta_\alpha}$\,.
Its irreducible decomposition can be expressed as
\begin{equation}\label{eq: M=irred}
\widetilde{M}_{\alpha\beta\sigma}={}^{(1)}\widetilde{M}_{\alpha\beta\sigma}+{}^{(2)}\widetilde{M}_{\alpha\beta\sigma}+\widetilde{\mathcal{S}}_{\alpha\beta\sigma}+\widetilde{m}_{\alpha\beta\sigma}\,,
\end{equation}
where
\begin{align*}
    &{}^{(1)}\widetilde{M}_{\alpha\beta\sigma}:=\frac{1}{4}\widetilde{a}_\alpha g_{\beta\sigma}\,,\\
    &{}^{(2)}\widetilde{M}_{\alpha\beta\sigma}:=\frac{1}{36}(g_{\beta\sigma}\delta_\alpha^\mu-2g_{\alpha\beta}\delta_\sigma^\mu-2g_{\alpha\sigma}\delta_\beta^\mu)(\widetilde{a}_\mu-4\widetilde{b}_\mu)\,,\\
    &\widetilde{\mathcal{S}}_{\alpha\beta\sigma}=\widetilde{M}_{(\alpha\beta\sigma)}-\frac{1}{18}(g_{\alpha\beta}\delta^\mu_\sigma+g_{\sigma\alpha}\delta^\mu_\beta+g_{\beta\sigma}\delta^\mu_\alpha)(\widetilde{a}_\mu+2\widetilde{b}_\mu)\,,\\
    &\widetilde{m}_{\alpha\beta\sigma}:=\widetilde{M}_{\alpha\beta\sigma}-{}^{(1)}\widetilde{M}_{\alpha\beta\sigma}-{}^{(2)}\widetilde{M}_{\alpha\beta\sigma}-\widetilde{\mathcal{S}}_{\alpha\beta\sigma}\,.
\end{align*}
All these tensors are symmetric in the last two indices and $\widetilde{\mathcal{S}}$ and $\widetilde{m}$ are traceless. Moreover, we have also the ciclicity condition analogous to \eqref{eq: t antisym=0}
\[\widetilde{m}_{\alpha\beta\sigma}+\widetilde{m}_{\sigma\alpha\beta}+\widetilde{m}_{\beta\sigma\alpha}=0\qquad\longrightarrow\qquad\widetilde{m}_{(\alpha\beta\sigma)}=0\]
We have obtained a decomposition of any $3$-tensor symmetric in its last two indices given by \eqref{eq: M=irred}. Conversely, we have that this decomposition completely characterizes the tensor:
\begin{remark}\mbox{}\\
The irreducible decomposition of a $3$-tensor symmetric in its last two indices is given by two $1$-forms $(a_\beta,b_\beta)$, a completely symmetric and traceless $3$-tensor $S_{\alpha\beta\sigma}$, and a $3$-tensor $m_{\alpha\beta\sigma}$ satisfying
\begin{equation}\label{eq: properties m}
\widetilde{m}_{\alpha\beta\sigma}=\widetilde{m}_{\alpha\sigma\beta}\qquad\qquad \tensor{\widetilde{m}}{^\alpha_\alpha_\sigma}=0=\tensor{\widetilde{m}}{_\alpha_\beta^\beta}\qquad\qquad\widetilde{m}_{\alpha\beta\sigma}+\widetilde{m}_{\sigma\alpha\beta}+\widetilde{m}_{\beta\sigma\alpha}=0
\end{equation}
\end{remark}
To prove that, notice that a $3$-tensor with this symmetry has $n\frac{n(n+1)}{2}$ components, 40 for $n=4$, and $(a_\beta,b_\beta,\mathcal{S}_{\alpha\beta\sigma},m_{\alpha\beta\sigma})$ has
\[n+n+\frac{1}{6}n(n-1)(n+4)+\frac{1}{3}n(n^2-4)\]
components, also 40 for $n=4$. Therefore, $(\widetilde{a}_\beta,\widetilde{b}_\beta,\widetilde{\mathcal{S}}_{\alpha\beta\sigma},\widetilde{m}_{\alpha\beta\sigma})$ is the irreducible decomposition of $\widetilde{M}_{\alpha\beta\sigma}$. 

\subsubsection*{Expressing the equation of motion in terms of the irreducible decompositions}
Once we have decomposed the torsion and nonmetricity, we rewrite  the equations of motion in terms of these irreducible components. Plugging equations \eqref{eq: Tor=irred} and \eqref{eq: M=irred} into the expression \eqref{eq: Q=Tor+M} of $Q$ in terms of the torsion and nonmetricity (recall that the torsion and the nonmetricity of the LC connection are zero) leads to
\begin{align}\begin{split}
    \tensor{Q}{^\alpha_\beta_\sigma}&=\frac{1}{3}(\widetilde{T}^\alpha g_{\beta\sigma}-\widetilde{T}_\sigma \delta^\alpha_\beta)+\frac{1}{36}\Big((2\widetilde{b}_\sigma-5 \widetilde{a}_\sigma) \delta^\alpha_\beta+ (2\widetilde{b}_\beta-5\widetilde{a}_\beta) \delta^\alpha_\sigma+(7 \widetilde{a}^\alpha -10 \widetilde{b}^\alpha) g_{\beta\sigma}\Big)\\
    &+\frac{1}{2}\tensor{\widetilde{\mathcal{A}}}{^\alpha_\beta_\sigma}-\frac{1}{2}\tensor{\widetilde{\mathcal{S}}}{^\alpha_\beta_\sigma}-\tensor{\widetilde{t}}{_\beta_\sigma^\alpha}+\tensor{\widetilde{m}}{^\alpha_\beta_\sigma}
\end{split}\label{eq: Q = T t a b S}
\end{align}

Meanwhile, plugging this expression into $\mathcal{E}^{(m)}$ (we do not write the volume to ease the notation) leads to
\begin{align}
\label{eq:movimiento}   \begin{split}
 &\tensor{\mathcal{E}}{^\alpha^\beta^\sigma}:=\frac{1}{\vol}\big(\mathcal{E}^{(m)}\big)^{\alpha\beta\sigma}=\frac{1}{\vol}\bigg(\mathcal{E}_{\textrm{P-CP}}^{(m)}-\frac{1}{\gamma}\mathcal{E}_{\textrm{H-CP}}^{(m)}\bigg)^{\!\alpha\beta\sigma} \\
    &=\left(\frac{2}{3}\widetilde{T}^\sigma+\frac{4}{9}\widetilde{a}^\sigma-\frac{7}{9}\widetilde{b}^\sigma\right)g^{\alpha\beta}-\left(\frac{2}{3}\widetilde{T}^\alpha+\frac{2}{9}\widetilde{a}^\alpha+\frac{1}{9}\widetilde{b}^\alpha\right)g^{\sigma\beta}-\frac{1}{18}(\widetilde{a}^\beta-4 \widetilde{b}^\beta)g^{\sigma\alpha} \\
    &
   -\tensor{\widetilde{\mathcal{A}}}{^\alpha^\beta^\sigma}+\tensor{\widetilde{\mathcal{S}}}{^\alpha^\beta^\sigma}-\tensor{\widetilde{t}}{^\beta^\sigma^\alpha}+\tensor{\widetilde{m}}{^\alpha^\beta^\sigma}\\
    &-\frac{1}{2\gamma}\left\{\frac{2}{3}(2\widetilde{T}_\mu+\widetilde{a}_\mu-\widetilde{b}_\mu)\delta^\sigma_\nu\delta^\alpha_\rho+(\tensor{\widetilde{\mathcal{A}}}{^\sigma_\mu_\nu}+\tensor{\widetilde{t}}{^\sigma_\mu_\nu})\delta^\alpha_\rho
    -\left(\tensor{\widetilde{\mathcal{A}}}{^\alpha_\mu_\nu}+\tensor{\widetilde{t}}{^\alpha_\mu_\nu}-2\tensor{\widetilde{m}}{_\nu_\mu^\alpha}\right)\delta^\sigma_\rho\right\}\tensor{\vol}{^\rho^\beta^\mu^\nu}
    \end{split}
\end{align}
Since there are three free indices, these are 64 equations (not all of them independent) in the variables $(\widetilde{T}_\beta,\widetilde{d}_\beta,\tensor{\widetilde{t}}{^\alpha_\beta_\sigma},\widetilde{a}_\beta,\widetilde{b}_\beta,\widetilde{\mathcal{S}}_{\alpha\beta\sigma},\widetilde{m}_{\alpha\beta\sigma})$ which, as we mentioned before, have 64 independent components in total.\vspace*{2ex}

\subsubsection*{Solving for $\widetilde{T}, \widetilde{d}, \widetilde{a}, \widetilde{b}$}
Consider the following system of equations
\begin{equation}
    \begin{array}{lcl}
\tensor{\mathcal{E}}{_\alpha^\alpha^\sigma}=0  &\phantom{espacioespacio} & 2\widetilde{T}^\sigma+\dfrac{3}{2}\widetilde{a}^\sigma-3\widetilde{b}^\sigma-\dfrac{1}{8\gamma}\widetilde{d}^\sigma=0\\
\tensor{\mathcal{E}}{^\sigma^\beta_\beta}=0 & \equiv & 2\widetilde{T}^\sigma+\dfrac{1}{2}\widetilde{a}^\sigma+\widetilde{b}^\sigma-\dfrac{1}{8\gamma}\widetilde{d}^\sigma=0\\
\vol_{\alpha\beta\sigma\mu}\tensor{\mathcal{E}}{^\alpha^\beta^\sigma}=0 & & \dfrac{\gamma}{4}\widetilde{d}_\mu+4\widetilde{T}_\mu+2\widetilde{a}_\mu-2\widetilde{b}_\mu=0\,,
\end{array}
\end{equation}
where we have used that  $\vol_{\alpha\beta\sigma\mu}\widetilde{t}^{\alpha\beta\sigma}=0$, which follows from \eqref{eq: t antisym=0}. The solutions to this system of linear equations are
\begin{equation}\label{eq: sol a,b,T,d}
    \widetilde{a}^\sigma=-8U^\sigma\quad\quad \widetilde{b}^\sigma=-2U^\sigma\quad\quad \widetilde{T}^\sigma=3U^\sigma\quad\quad \widetilde{d}^\sigma=0
\end{equation}
for an arbitrary vector field $U^\sigma$ on $M$.\vspace*{2ex}

\subsubsection*{Solving for $\widetilde{\mathcal{S}}$}
Plugging the solutions \eqref{eq: sol a,b,T,d} into \eqref{eq:movimiento} leads to
\begin{equation}\label{eq: E half shell}
    \tensor{\mathcal{E}}{^\alpha^\beta^\sigma}=\tensor{\widetilde{\mathcal{S}}}{^\alpha^\beta^\sigma}-\tensor{\widetilde{t}}{^\beta^\sigma^\alpha}+\tensor{\widetilde{m}}{^\alpha^\beta^\sigma}
-\frac{1}{2\gamma}\left\{\tensor{\vol}{^\alpha^\beta^\mu^\nu}\tensor{\widetilde{t}}{^\sigma_\mu_\nu}
-\tensor{\vol}{^\sigma^\beta^\mu^\nu}\left(\tensor{\widetilde{t}}{^\alpha_\mu_\nu}-2\tensor{\widetilde{m}}{_\nu_\mu^\alpha}\right)\right\}
\end{equation}
By completely symmetrising this expression, all terms vanish except for the first one, so
\begin{equation}\label{sol:S}
    \widetilde{\mathcal{S}}^{\alpha\beta\sigma}=0
\end{equation}

\subsubsection*{Solving for $\widetilde{m}$}
Plugging \eqref{sol:S} into \eqref{eq: E half shell}, symmetrizing in $(\alpha,\sigma)$, using the cyclicity of $\widetilde{m}$, and imposing it to be zero leads to
\begin{equation}
    \tensor{\widetilde{m}}{^\alpha^\beta^\sigma}+
     \frac{1}{\gamma}\left(\tensor{\vol}{^\alpha^\beta_\mu_\nu}\tensor{\widetilde{m}}{^\mu^\nu^\sigma}+\tensor{\vol}{^\alpha^\sigma_\mu_\nu}\tensor{\widetilde{m}}{^\mu^\nu^\beta}\right)=0
     \end{equation}
or equivalently
\begin{equation}
    \tensor{\widetilde{m}}{^\alpha^\beta^\sigma}=\tensor{\mathcal{M}}{^\alpha^\beta^\sigma_\mu_\nu_\kappa}\tensor{\widetilde{m}}{^\mu^\nu^\kappa}\qquad\text{where}\qquad
\tensor{\mathcal{M}}{^\alpha^\beta^\sigma_\mu_\nu_\kappa}:=-\frac{1}{\gamma}\Big(\tensor{\vol}{^\alpha^\beta_\mu_\nu}\delta^\sigma_\kappa+\tensor{\vol}{^\alpha^\sigma_\mu_\nu}\delta^\beta_\kappa\Big)
\end{equation}
Applying this equation recursively leads to
\begin{equation}\label{sol:m}
\tensor{\widetilde{m}}{^\alpha^\beta^\sigma}=\tensor{\mathcal{M}}{^\alpha^\beta^\sigma _\mu_\nu_\kappa}\tensor{\widetilde{m}}{^\mu^\nu^\kappa}=\tensor{\mathcal{M}}{^\alpha^\beta^\sigma_\mu_\nu_\kappa}\tensor{\mathcal{M}}{^\mu^\nu^\kappa_\xi_\rho_\tau}\tensor{\widetilde{m}}{^\xi^\rho^\tau}\overset{\eqref{eq: properties m}}{=}-\frac{9}{\gamma^2}\tensor{\widetilde{m}}{^\alpha^\beta^\sigma}\quad\longrightarrow\quad \tensor{\widetilde{m}}{^\alpha^\beta^\sigma}=0
\end{equation}

\subsubsection*{Solving for $\widetilde{t}$} Plugging the solutions \eqref{sol:S} and \eqref{sol:m} into \eqref{eq: E half shell} and imposing it to be zero leads to 
\begin{equation}
    \tensor{\widetilde{t}}{^\alpha^\beta^\sigma}=\tensor{\mathcal{T}}{^\alpha^\beta^\sigma_\mu_\nu_\kappa}\tensor{\widetilde{t}}{^\mu^\nu^\kappa}\qquad\text{where}\qquad
     \tensor{\mathcal{T}}{^\alpha^\beta^\sigma_\mu_\nu_\kappa}:=\frac{1}{2\gamma}\Big(\tensor{\vol}{^\alpha^\beta_\mu_\nu}\delta^\sigma_\kappa-\tensor{\vol}{^\alpha^\sigma_\mu_\nu}\delta^\beta_\kappa\Big)
     \end{equation}
Applying this equation recursively leads to
\begin{equation}
\tensor{\widetilde{t}}{^\alpha^\beta^\sigma}=\tensor{\mathcal{T}}{^\alpha^\beta^\sigma_\mu_\nu_\kappa}\tensor{\widetilde{t}}{^\mu^\nu^\kappa}=\tensor{\mathcal{T}}{^\alpha^\beta^\sigma_\mu_\nu_\kappa}\tensor{\mathcal{T}}{^\mu^\nu^\kappa_\xi_\rho_\tau}\tensor{\widetilde{t}}{^\xi^\rho^\tau}\overset{\eqref{eq: properties t}}{=}-\frac{1}{\gamma^2}\tensor{\widetilde{t}}{^\alpha^\beta^\sigma}\quad\longrightarrow\quad \tensor{\widetilde{t}}{^\alpha^\beta^\sigma}=0
\end{equation}

\subsubsection*{Final space of solutions}
By solving the equation of motion $\mathcal{E}^{(m)}=0$, we have obtained for the independent variables $(\widetilde{T}_\beta,\widetilde{d}_\beta,\tensor{\widetilde{t}}{^\alpha_\beta_\sigma},\widetilde{a}_\beta,\widetilde{b}_\beta,\widetilde{\mathcal{S}}_{\alpha\beta\sigma},\widetilde{m}_{\alpha\beta\sigma})$ the following necessary conditions:
\begin{equation}
    \widetilde{a}^\sigma=-8U^\sigma\quad\quad \widetilde{b}^\sigma=-2U^\sigma\quad\quad \widetilde{T}^\sigma=3U^\sigma\quad\quad \widetilde{d}^\sigma=0\quad\quad\widetilde{t}^{\alpha\beta\sigma}=\widetilde{\mathcal{S}}^{\alpha\beta\gamma}=\widetilde{m}^{\alpha\beta\gamma}=0\label{eq: sol a b T}
\end{equation}
for an arbitrary $U^\sigma$. Plugging these into \eqref{eq:movimiento} shows they are also sufficient, so the first equation of motion is completely solved. Moreover, if we insert \eqref{eq: sol a b T} into \eqref{eq: Q = T t a b S}, the expression of $Q$ over the space of solutions is given by
\begin{equation}\label{eq: Q=delta U}
\tensor{Q}{^\alpha_\beta_\sigma}=\delta^\alpha_\sigma U_\beta
\end{equation}
Substituting back this result into the other equation of motion, it is straightforward to check that it vanishes, hence only the Einstein equations remain:
\[0=\mathfrak{E}^{(m)}:=\mathfrak{E}_{\textrm{EH}}^{(m)} +\mathfrak{E}_{\textrm{P-CP}}^{(m)}-\frac{1}{\gamma}\mathfrak{E}_{\textrm{H-CP}}^{(m)} 
 \overset{\eqref{eq: Q=delta U}}{=} \mathfrak{E}_{\textrm{EH}}^{(m)} \]
 This means that $(g,Q)$ is a solution for metric-HMS action if and only if $\tensor{Q}{^\alpha_\beta_\sigma}=\delta^\alpha_\sigma U_\beta$ and $g$ satisfies the Einstein equations:
\begin{equation}\label{eq: Sol S m HMS}
\comentario\tikzmarkin{equivSolprima}(0.2,-0.24)(-0.2,0.55)
	\tikzmarkin{equivSol}(0.2,-0.19)(-0.2,0.5)
\mathrm{Sol}^{(m)}_{\mathrm{HMS}}=\{(g_{\alpha\beta},\delta^\alpha_\gamma U_{\!\beta})\ |\ g\in\mathrm{Sol}^{(m)}_{\mathrm{GR}},\ U_{\!\beta}\text{ arbitrary}\}\overset{\text{\cite{CPSPT}}}{=}\mathrm{Sol}^{(m)}_{\mathrm{PT}}	\tikzmarkend{equivSol}
	\tikzmarkend{equivSolprima}
\end{equation}
This result proves that the metric sector of the metric-HMS theory is equivalent to the metric-EH theory as explained in \cite{CPSGR,CPSPT}. We have the following on shell identities:
   \begin{equation}\label{eq: tilde=LC+}
   \begin{array}{ll}
       \mathrm{\widetilde{R}}\tensor{\mathrm{iem}}{^\alpha_\beta_\mu_\nu}=\mathrm{\LC{R}}\tensor{\mathrm{iem}}{^\alpha_\beta_\mu_\nu}+g^\alpha_\beta(\d U)_{\mu\nu}\,,\\[0.3ex]
        \mathrm{\widetilde{R}ic}_{\beta\nu}=\mathrm{\LC{R}ic}_{\beta\nu}+(\d U)_{\beta\nu}\,,&  \widetilde{R}=\LC{R}\,,\\
        \widetilde{K}_{\overline{\alpha}\overline{\beta}}=\LC{K}_{\overline{\alpha}\overline{\beta}}-\dfrac{1}{2}(U\wedge\nu)_{\overline{\alpha}\overline{\beta}}\,,&  \widetilde{K}=\LC{K}\,,\\ 
        \widetilde{M}_{\alpha\beta\gamma}=-2 g_{\beta\gamma}U_\alpha\,,& \tensor{\widetilde{\mathrm{T}}\mathrm{or}}{^\gamma_\alpha_\beta}=\delta^\gamma_\beta U_\alpha-\delta^\gamma_\alpha U_\beta\,,\\
        \hat{L}^{(m)}_{\textrm{P-CP}}=0\,,&\hat{L}^{(m)}_{\textrm{H-CP}}=0\,.\end{array}
    \end{equation}

\subsection{Presymplectic form}
     Taking the $\dd$-exterior derivative of \eqref{eq: Theta CP} we obtain:
     \[\dd\Theta^{(m)}_{\mathrm{HMS}}=\dd\Theta^{(m)}_{\mathrm{PT}}={\dd\Theta^{(m)}_{\mathrm{EH}
     }}
     \]
     Thus, the metric-HMS presymplectic form $\OOmega^{\mathrm{HMS}}_{(m)}$ defined over the space of solutions $\mathrm{Sol}^{(m)}_{\mathrm{HMS}}$ is the same as the one of metric-Palatini $\OOmega^{\mathrm{HMS}}_{(m)}$  which, in turn, has the same functional form as the metric-EH presymplectic form $\OOmega^{\mathrm{EH}}_{(m)}$ (see \cite{CPSPT,CPSGR}). In fact, if we define the projection $\pi_{(m)}(g,Q)=g$, the presymplectic forms canonically associated to the three actions can be put into correspondence. 
     \begin{equation}\label{eq: O_PT(m)=pi Omega_GR(m)}
\comentario\tikzmarkin{equivOmegaprima}(0.2,-0.29)(-0.2,0.52)
	\tikzmarkin{equivOmega}(0.2,-0.24)(-0.2,0.47)
         \OOmega^{\mathrm{HMS}}_{(m)}=\OOmega^{\mathrm{PT}}_{(m)}=\pi_{(m)}^*\OOmega^{\mathrm{{EH}}}_{(m)}
\tikzmarkend{equivOmega}
	\tikzmarkend{equivOmegaprima}
\end{equation}

\section{METRIC-HMS WITH BOUNDARY}

\subsection{The action}
In this section, we consider a manifold with non-empty boundary. The action over the space of fields $\mathcal{F}^{(m)}_{\textrm{HMS}}:=\{(g,Q)|\jmath^*\!g\text{ Lorentzian}\}$ is given by
\[
\mathbb{S}^{(m)}_{\textrm{HMS}}:=\int_M L^{(m)}_{\textrm{HMS}}-\int_{\partial_L M}\overline{\ell}^{(m)}_{\textrm{HMS}}\,,
\]
with the same bulk Lagrangian \eqref{eq: L=L+L+d something} and a boundary Lagrangian
     \begin{equation}\label{eq: Lagrangians metric}
\definicion\tikzmarkin{Lagrangianpairprima}(0.25,-0.44)(-0.25,0.65)
	\tikzmarkin{Lagrangianpair}(0.2,-0.44)(-0.2,0.65)
         L^{(m)}_{\textrm{HMS}}:=L^{(m)}_{\textrm{PT}}-\frac{1}{2\gamma}\vol^{\alpha\beta\mu\nu}\widetilde{\mathrm{R}}\mathrm{iem}_{\alpha\beta\mu\nu}\vol\qquad\overline{\ell}^{(m)}_{\textrm{HMS}}:=\overline{\ell}^{(m)}_{\textrm{PT}}+\frac{1}{\gamma}\jmath^*\iota_{\vec{q}}\vol
\tikzmarkend{Lagrangianpair}
	\tikzmarkend{Lagrangianpairprima}
\end{equation} 
where we recall that $q^\mu=\vol^{\mu\alpha\beta\nu}Q_{\alpha\beta\nu}$. The first term of the boundary Lagrangian is the generalized Hawking-Gibbons-York term introduced by Obukhov in \cite{Obukohv1987} (see also \cite{CPSPT})
\[\overline{\ell}^{(m)}_{\mathrm{PT}}(g,Q):=-2\widetilde{K} \, \vol_{\overline{g}}\,,\qquad\qquad\overline{\ell}^{(m)}_{\mathrm{GHY}}(g):=-2\LC{K} \, \vol_{\overline{g}}\,,\]
while the second term is a new one introduced to cancel the exact terms coming from the variation of the bulk Lagrangian. It is interesting to note that, using equation \eqref{eq: K=K+Q}, the HMS boundary term can be written as
\[\overline{\ell}^{(m)}_{\textrm{HMS}}(g, Q)=\overline{\ell}^{(m)}_{\textrm{GHY}}(g)+\jmath^*\iota_{\vec{A} - \vec{C}+\vec{q}/\gamma}\vol\,,\qquad  \quad 
     \]
which, in view of equation \eqref{eq: L=L+L+d something}, shows why this is a natural choice of boundary Lagrangian.

\subsection{Variations}
From the computations of the previous section, we obtain the same result in the bulk
\begin{align*}
&\dd L^{(m)}_{\textrm{HMS}}=(\mathfrak{E}^{(m)})^{\alpha \beta} \dd g_{\alpha \beta} +\tensor{(\mathcal{E}^{(m)})}{_\alpha^\beta^\sigma} \dd \tensor{Q}{^\alpha_\beta_\sigma}+ \d\Big(\Theta^{(m)}_{\textrm{EH}}+ \dd(\iota_{\vec{A} - \vec{C}+\vec{q}/\gamma}\vol)\Big)
\end{align*}
Now, following the CPS algorithm \cite{CPS}, we compute on the lateral boundary
\begin{align*}
\dd\overline{\ell}^{(m)}_{\textrm{HMS}}-\jmath^*\Theta_{\textrm{HMS}}^{(m)}&=\dd\overline{\ell}^{(m)}_{\textrm{GHY}}+\dd\jmath^*\iota_{\vec{A} - \vec{C}+\vec{q}/\gamma}\vol-\jmath^*\Theta^{(m)}_{\textrm{EH}}-\jmath^* \dd(\iota_{\vec{A} - \vec{C}+\vec{q}/\gamma}\vol)\\
&=\dd\overline{\ell}^{(m)}_{\textrm{GHY}}-\jmath^*\Theta^{(m)}_{\textrm{EH}}=\overline{b}_{(m)}^{\overline{\alpha}\overline{\beta}}\dd\overline{g}_{\overline{\alpha}\overline{\beta}}-\d\overline{\theta}^{(m)}_{\mathrm{HMS}}\,,
\end{align*}
where in the last equality we have the usual quantities of EH (see for instance \cite{CPSGR}) 
\begin{align}
   &\overline{b}_{(m)}^{\overline{\alpha}\overline{\beta}}(g)=\left(\LC{K}^{\overline{\alpha}\overline{\beta}}-\LC{K}\overline{g}^{\overline{\alpha}\overline{\beta}}\right)\vol_{\overline{g}}\,,\nonumber\\
   &\overline{\theta}^{(m)}_{\mathrm{HMS}}:=\overline{\theta}^{(m)}_{\mathrm{EH}}=\iota_{\bar{V}}\vol_{\bar{g}}\,,\label{theta_HMS_m}\\
   &\overline{V}_{\!\overline{\alpha}}=-\jmath^\alpha_{\overline{\alpha}}\nu^\beta\dd g_{\alpha\beta}\,.\nonumber
\end{align}

\subsection{Space of solutions}
Although the manifold we are considering has a boundary term, the boundary equations only involve the metric, and hence we recover the same result \eqref{eq: Sol S m HMS}. This implies that the boundary only plays a role in the metric sector of the solution space $\mathrm{Sol}^{(m)}_{\mathrm{EH}}$ (discussed in detail in \cite{CPSGR}, where both Dirichlet and Neumann boundary conditions were considered). The same techniques used here apply if one considers other boundary terms to impose different boundary conditions.

\subsection{Presymplectic form}
We have seen that $\overline{\theta}^{(m)}_{\mathrm{HMS}}=\overline{\theta}^{(m)}_{\mathrm{PT}}=\overline{\theta}^{(m)}_{\mathrm{EH}}$. This together with \eqref{eq: Theta CP} leads once again to \eqref{eq: O_PT(m)=pi Omega_GR(m)}.

\section{TETRAD-HMS WITH BOUNDARY}\label{sec: tetrad no boundary}
 
\subsection{The action}
The tetrad-HMS action is defined as $\SS_{\textrm{HMS}}^{(t)}:=\SS_{\textrm{HMS}}^{(m)}\circ\Phi_{\mathrm{HMS}}$ where 
\[\Phi_{\mathrm{HMS}}(e,\widetilde{\omega}):=\Phi_{\mathrm{PT}}(e,\widetilde{\omega})\overset{\text{\cite{CPSPT}}}{=}\Big(\eta_{IJ}e_\mu^I e_\nu^J,E^\beta_Ke^J_\alpha\big(\tensor{\widetilde{\omega}}{_\mu^K_J}-\tensor{\LC{\omega}}{_\mu^K_J}\big)\Big)=(g_{\mu\nu},\tensor{Q}{^\beta_\mu_\alpha})\,,\]
As usual, $e^I_\alpha$ are the tetrad $1$-forms, $E_I^\alpha$ the dual cotetrad vector fields, $\eta_{IJ}$ the internal Lorentz metric, $\widetilde{\omega}^{IJ}$ is the generic $1$-form connection (no symmetries in the internal indices), and $\LC{\omega}^{IJ}$ is the Levi-Civita $1$-form connection associated with the metric $g_{\mu\nu}=\eta_{IJ}e^I_\mu e^J_\nu$. \vspace*{2ex}

$\Phi_{\mathrm{HMS}}$ is almost a change of variables from tetrads and spin connections $(e,\widetilde{\omega})\in\mathcal{F}^{(t)}_{\textrm{HMS}}$ to metrics and $3$-tensors $(g,Q)\in\mathcal{F}^{(m)}_{\textrm{HMS}}$: it is surjective but not injective. In fact, \begin{equation}\label{eq: Phi lambda-inv} \Phi_{\mathrm{HMS}}(e,\widetilde{\omega})=\Phi_{\mathrm{HMS}}(e',\widetilde{\omega}')\, \longleftrightarrow\, \exists\text{\,local }\Psi\in SO(1,3)\  \ \begin{array}{|l}
 e'_I=\tensor{\Psi}{_I^J}e_J\\
 \widetilde{\omega}_I ^{'\,J}=\tensor{\Psi}{_I^K}\widetilde{\omega}\tensor{}{_K^L}\tensor{\Psi}{^J_L}+\tensor{\Psi}{_I^K}\d\tensor{\Psi}{^J_K}
\end{array}
\end{equation}
Moreover, it is not too hard to check that $\dd\Phi_{\mathrm{HMS}}$ is surjective.
\begin{remark}\label{remark: dd Phi derivatives}\mbox{}\\
Naively, one might think that the fact that $\Phi_{\mathrm{HMS}}$ is almost a change of variables implies that the results derived in the metric formalism can be easily transferred into the tetrad formalism. However, this is not the case since $\dd\Phi_{\mathrm{HMS}}$ involves derivatives of $\dd e^I_\alpha$ (due to the variation of $\tensor{\LC{\omega}}{_\mu^K_I}:= e^{K}_{\alpha}\LC{\nabla}_{\!\mu}E^{\alpha}_{I}$). These derivatives will appear in the computation of the symplectic potentials and charges.
\end{remark}

From the aforementioned definition of the tetrad-HMS action, we choose the following representatives as the tetrad-HMS Lagrangians
\begin{align}\label{eq: L HMS}
    &L^{(t)}_{\mathrm{HMS}}(e, \widetilde{\omega}) :=L^{(m)}_{\mathrm{HMS}}\circ\Phi_{\mathrm{HMS}}(e, \widetilde{\omega})\,,&\overline{\ell}^{(t)}_{\mathrm{HMS}}(e, \widetilde{\omega}) :=\overline{\ell}^{(m)}_{\mathrm{HMS}}\circ\Phi_{\mathrm{HMS}}(e, \widetilde{\omega})\,.\
\end{align}
In this case, it is not really useful to write the Lagrangian as the sum of the EH-term and some coupling terms (in analogy with \eqref{eq: L=L+L+d something}). It is better to split the $1$-form connection in its antisymmetric and symmetric parts in its internal indices,
\begin{equation*}
    \widetilde{\omega}_{IJ} = \widehat{\omega}_{IJ} + S_{IJ}\,,
\end{equation*}
and consider the equivalent variables $(e,\widehat{\omega},S)$. Following \cite{CPSGR,CPSPT}, one obtains the following explicit expressions for these Lagrangians
\begin{align*}
    \comentario\tikzmarkin{LagHMS}(0.7,-0.45)(-0.2,0.65)
    \tikzmarkin{LagHMSprima}(0.7,-0.4)(-0.2,0.6)
    &L_{\mathrm{HMS}}^{(t)}(e,\widehat{\omega},S)=\frac{1}{2}H_{IJKL}\Big( \widehat{F}^{IJ}- \frac{\Lambda}{6}e^{I} \wedge e^{J}+\tensor{S}{^I_M}\wedge S^{MJ} \Big) \wedge e^K\wedge e^L\\
    &\overline{\ell}^{(t)}_{\mathrm{HMS}}(e, \widehat{\omega})= -\frac{1}{2}H_{IJKL}\left(2 N^{I}\d N^{J} - \overline{\widehat{\omega}}{}^{IJ}\right)\wedge \overline{e}^K\wedge \overline{e}^{L}-\frac{1}{\gamma}\d \overline{e}_I\wedge\overline{e}^I
\tikzmarkend{LagHMS}
\tikzmarkend{LagHMSprima}
\end{align*}
where $\tensor{H}{_I_J^K^L}=\tensor{\varepsilon}{_I_J^K^L}+\frac{1}{\gamma}(\delta_I^K\delta_J^L-\delta_I^L\delta_J^K)$, $\widehat{F}_{IJ} = \d \widehat{\omega}_{IJ} + \widehat{\omega}_{IK}\wedge \tensor{\widehat{\omega}}{^K_J}$, $\overline{e}^I:=\jmath^*e^I$, $\overline{\widehat{\omega}}{}^{IJ}:=\jmath^*\widehat{\omega}^{IJ}$, and $N^I=\nu^\alpha e^I_\alpha$ . Notice that if we set $S=0$ we recover the Holst Lagrangian on the bulk
\[L_{\mathrm{HMS}}^{(t)}=L_{\mathrm{Holst}}^{(t)}+\frac{1}{2}H_{IJKL}\tensor{S}{^I_M}\wedge S^{MJ}  \wedge e^K\wedge e^L\]
while the boundary term is the one defined in \cite{bodendorfer2013imaginary} (although in that reference it is derived by working from the beginning with a Lorentz connection i.e. $S=0$). Notice in particular that  the tetrad-HMS action can be interpreted as the generalization of the Holst action for $GL(4)$ connections.

\subsection{Variations}
Computing the variations, one easily obtains
\begin{align*}
    & \dd L^{(t)}_{\mathrm{HMS}} = \mathfrak{E}^{(t)}_{L}\wedge \dd e^{L} + \mathcal{E}^{(t)}_{KL}\wedge \dd \widehat{\omega}^{KL}+ \mathscr{E}^{(t)}_{JM}\wedge \dd S^{JM}  + \d \Theta^{(t)}_{\mathrm{HMS}}\,, \\
   & \dd \overline{\ell}^{(t)}_{\mathrm{HMS}} - \jmath^{*}\Theta^{(t)}_{\mathrm{HMS}} = \overline{b}^{(t)}_{I}\wedge \dd \overline{e}^{I} - \d \overline{\theta}^{(t)}_{\mathrm{HMS}}\,,
\end{align*}
where the Euler-Lagrange equations are
\begin{align}\begin{split}\label{eq: eq of motions tetrad}
    \mathfrak{E}^{(t)}_{L} &:= H_{IJKL}\Big( \widehat{F}^{IJ} +\tensor{S}{^I_M}\wedge S^{MJ}- \frac{\Lambda}{3} e^{I} \wedge e^{J}\Big) \wedge e^{K}\,,\\
    \mathcal{E}^{(t)}_{KL} &:= -\frac{1}{2}\widehat{D}(H_{IJKL}e^{I}\wedge e^{J})\,,\\
    \mathscr{E}^{(t)}_{JM}&:= \frac{1}{2}\big( H_{IKLJ}\delta_M^R + H_{IKLM}\delta_J^R\big)\tensor{S}{_R ^I}\wedge e^{K} \wedge e^{L}\,,\\
     \overline{b}^{(t)}_{I} &:=\epsilon_{IJKL}(2N^{K}\d N^{L} - \overline{\widehat{\omega}}{}^{KL})\wedge \overline{e}^{J} +2\epsilon_{MJKL}N^L(\iota_{\overline{E}^J}\d\overline{e}^K)\wedge\overline{e}^MN_I-\frac{2}{\gamma}\widehat{D}\overline{e}_I\,,\end{split}
\end{align}
here $\widehat{\mathcal{D}}\alpha_I=\mathrm{d}\alpha_I+\tensor{\widehat{\omega}}{_I^J}\wedge\alpha_J$. We take the symplectic potentials
\begin{align}\label{eq: Theta tetrad HMS}
    &\Theta^{(t)}_{\mathrm{HMS}} := \frac{1}{2}H_{IJKL}e^{I}\wedge e^{J} \wedge \dd \widehat{\omega}^{KL}\,,
    &\overline{\theta}^{(t)}_{\mathrm{HMS}} := \epsilon_{IJKL}\overline{e}^{I}\wedge \overline{e}^{J}  \wedge N^{K} \dd N^{L}-\frac{1}{\gamma}\overline{e}^I\wedge\dd\overline{e}_I\,.
\end{align}

\subsection{Space of solutions}
One way to obtain the space of solutions is to prove that $\dd\Phi_{\mathrm{PT}}$ is surjective. Then, because 
\begin{equation}\label{eq: dd SS = chain rule}
    \dd_{(e,\widetilde{\omega})}\SS_{\textrm{HMS}}^{(t)}=\dd_{(e,\widetilde{\omega})}(\SS_{\textrm{HMS}}^{(m)}\circ\Phi_{\mathrm{PT}})=\dd_{\Phi_{\mathrm{PT}}(e,\widetilde{\omega})}\SS_{\textrm{HMS}}^{(m)}\circ\dd_{(e,\widetilde{\omega})}\Phi_{\mathrm{PT}}
\end{equation}
we have
\begin{equation}\label{eq: Sol S t HMS}
\comentario\tikzmarkin{equivSolHMSprima}(0.2,-0.24)(-0.2,0.6)
	\tikzmarkin{equivSolHMS}(0.2,-0.19)(-0.2,0.55)
\Sol_{\textrm{HMS}}^{(t)}\overset{\eqref{eq: dd SS = chain rule}}{=}\Phi_{\mathrm{PT}}^{-1}\Sol_{\textrm{HMS}}^{(m)}\overset{\eqref{eq: Sol S m HMS}}{=}\Phi_{\mathrm{PT}}^{-1}\Sol_{\textrm{PT}}^{(m)}\overset{\text{\cite{CPSPT}}}{=}\Sol_{\textrm{PT}}^{(t)}	\tikzmarkend{equivSolHMS}
	\tikzmarkend{equivSolHMSprima}
\end{equation}
Although it is not too hard to prove that $\dd\Phi_{\mathrm{PT}}$ is surjective, here we will use the techniques of section \ref{subsection: space solution metric no boundary} to solve $\mathscr{E}^{(t)}_{JM}=0$. To that purpose let us first expand $S_{IJ}=S_{MIJ}e^M$ with $S_{MIJ}=S_{MJI}$ to rewrite the third equation in \eqref{eq: eq of motions tetrad} in the form
\[
\epsilon_{KLNP}\tensor{H}{^{IKL}_{(M}}\tensor{S}{^N_{J)I}}=0\,,
\]
which is equivalent to
\begin{equation}\label{equation_S-expanded}
-2\tensor{S}{_{I(M}^I}\tensor{\delta}{_{J)}^P}+2\tensor{S}{_{(MJ)}^P}-\frac{1}{\gamma}S_{NL(M}\tensor{\epsilon}{_{J)}^{NLP}}=0\,.    
\end{equation}
We use now the irreducible decomposition \eqref{eq: M=irred} (notice that we are allowed to do that despite the different ``nature of the indices'') to write
\begin{equation}\label{eq: S=irred}
S_{IJK}={}^{(1)}S_{IJK}+{}^{(2)}S_{IJK}+\Sigma_{IJK}+\sigma_{IJK}
\end{equation}
with
\begin{align*}
    &{}^{(1)}S_{IJK}:=\frac{1}{4}A_I \eta_{JK}\,,\\
    &{}^{(2)}S_{IJK}:=\frac{1}{36}(\eta_{JK}\delta_I^L-2\eta_{IJ}\delta_K^L-2\eta_{IK}\delta_J^L)(A_L-4B_L)\,,\\
    &\Sigma_{IJK}=S_{(IJK)}-\frac{1}{18}(\eta_{IJ}\delta^L_K+\eta_{KI}\delta^L_J+\eta_{JK}\delta^L_I)(A_L+2B_L)\,,\\
    &\sigma_{IJK}:=S_{IJK}-{}^{(1)}S_{IJK}-{}^{(2)}S_{IJK}-\Sigma_{IJK}\,,
\end{align*}
where all these tensors are symmetric in the last two indices, $\Sigma_{(IJK)}=\Sigma_{IJK}$, $\Sigma_{IJK}$ and $\sigma_{IJK}$ are traceless and, finally, $\sigma_{IJK}+\sigma_{KIJ}+\sigma_{JKI}=0$. 
We now solve \eqref{eq: S=irred} in steps. First, by contracting it with $\delta^J_P$ we get
\[\tensor{S}{_{MI}^I}-4\tensor{S}{_{IM}^I}=0\qquad\longleftrightarrow\qquad A_M-4B_M=0\,.\]
By symmetrizing \eqref{eq: S=irred} in the indices $MJP$ we find
\[
S_{(JMP)}-\tensor{S}{_{I(M}}^I\eta_{JP)}=0\qquad\longleftrightarrow\qquad \Sigma_{JMP}+\frac{1}{6}(A_{(J}-4B_{(J})\eta_{MP)}=0\,.
\]
We then conclude that ${}^{(2)}S_{IJK}=0$ and $\Sigma_{IJK}=0$, so that $S_{IJK}=\frac{1}{4}A_I\eta_{JK}+\sigma_{IJK}$ and \eqref{eq: S=irred} becomes  
\[
\sigma_{PJM}=\tensor{\mathcal{N}}{_{PJM}^{TUV}}\sigma_{TUV}\qquad\qquad
\tensor{\mathcal{N}}{_{PJM}^{TUV}}:=\frac{1}{2\gamma}(\delta_M^U\tensor{\epsilon}{^T_J^V_P}+\delta_J^U\tensor{\epsilon}{^T_M^V_P})\,.
\]
Now,
\[
\sigma_{PJM}=\tensor{\mathcal{N}}{_{PJM}^{TUV}}\sigma_{TUV}=\tensor{\mathcal{N}}{_{PJM}^{TUV}}\tensor{\mathcal{N}}{_{TUV}^{ABC}}\sigma_{ABC}=-\frac{3}{\gamma^2}\sigma_{PJM}\,,
\]
as a consequence of the tracelessness of $\sigma_{IJK}$ and its ciclicity. We then conclude $\sigma_{PJM}=0$ for all $\gamma\in\mathbb{R}$, and the general solution for $S_{IJ}$ has the form
\[
S_{IJ}=\eta_{IJ}U_Ke^K
\]
with $U_{\!K}$ arbitrary. Plugging this solution into $\mathcal{E}^{(t)}_{KL}=0$ of \eqref{eq: eq of motions tetrad} removes the dependence in $S$ and the system becomes the ones studied in \cite{concise,unimodular}, where we found the solution $\widehat{\omega}_{IJ}=\LC{\omega}_{IJ}$. Finally,  once we plug the solutions for $S$ and $\widetilde{\omega}$, $e^I$ has to satisfy the Einstein equation coming from $\mathfrak{E}^{(t)}_{L}$.

\subsection{Presymplectic form}
From \eqref{eq: Theta tetrad HMS} and \cite{CPSPT,CPSGR}, we have
\begin{align}\label{eq: Theta tetrad HMS-PT}
    &\Theta^{(t)}_{\mathrm{HMS}}=\Theta^{(t)}_{\mathrm{PT}}+ \frac{1}{\gamma}e_I\wedge e_J \wedge \dd \widehat{\omega}^{IJ}\,,
    &\overline{\theta}^{(t)}_{\mathrm{HMS}}=\overline{\theta}^{(t)}_{\mathrm{PT}}-\frac{1}{\gamma}\overline{e}^I\wedge\dd\overline{e}_I\,.
\end{align}
Alternatively, defining the contorsion $C^{IJ}:=\widehat{\omega}^{IJ}-\LC{\omega}^{IJ}$, we can write
\begin{align*}
    \Theta^{(t)}_{\mathrm{HMS}}&=\Theta^{(t)}_{\mathrm{PT}}+ \frac{1}{\gamma}e_I\wedge e_J \wedge \dd C^{IJ}+\frac{1}{\gamma}e_I\wedge e_J\wedge\dd\LC{\omega}^{IJ}=\\
    &=\Theta^{(t)}_{\mathrm{PT}}+ \frac{1}{\gamma}e_I\wedge e_J \wedge \dd C^{IJ}-\frac{1}{\gamma}\d(e_I\wedge \dd e^I)\,.
\end{align*}
The last equality follows from the expression of $\LC{\omega}$ in terms of $\LC{D}$ and the fact that $\LC{D}e^I=0$ (since the connection is the LC one, there is no torsion). Gathering the previous equations and using the relative bicomplex framework \cite{CPS}, we obtain
\begin{align}
    \comentario\tikzmarkin{Tetradseq}(0.2,-0.5)(-0.2,0.75)
    \tikzmarkin{Tetradseqprima}(0.2,-0.45)(-0.2,0.7)
    \Big(\Theta^{(t)}_{\mathrm{HMS}},\overline{\theta}^{(t)}_{\mathrm{HMS}}\Big)=\Big(\Theta^{(t)}_{\mathrm{PT}},\overline{\theta}^{(t)}_{\mathrm{PT}}\Big)+ \frac{1}{\gamma}\Big(e_I\wedge e_J \wedge \dd C^{IJ},0\Big)-\frac{1}{\gamma}\underline{\d}\big(e_I\wedge \dd e^I,0\big)
\tikzmarkend{Tetradseq}
\tikzmarkend{Tetradseqprima}
\end{align}
Notice that, off-shell, the HMS and Palatini symplectic potentials are not equal in the relative cohomology due do the term involving $\dd C^{IJ}$. Moreover, we see that the relative cohomology class of the HMS symplectic potentials depends on $\gamma$ while in the Palatini case, of course, it does not. However, $C=0$ on-shell and we obtain, as in the metric case, the expected equivalence over the space of solutions and the independence of $\gamma$.

\section{CHARGES IN THE METRIC FORMALISM}

\subsection{The HMS and Palatini Lagrangian pairs are not equal in relative cohomology}\label{subsection: pairs cohomology metric}
The space of solutions and the symplectic structure of metric-HMS is the same as in metric-Palatini. However, the Lagrangians are not the same at all. In fact, they are not even in the same relative cohomological class so, in principle, the associated charges computed with the CPS algorithm may differ. Indeed, from our previous computations we have that
\begin{align*}
    &\dd(L^{(m)}_{\textrm{HMS}}-L^{(m)}_{\textrm{PT}})=-\frac{1}{\gamma}\Big(\mathfrak{E}_{\textrm{H-CP}}^{(m)}\Big)^{\alpha \beta} \dd g_{\alpha \beta} -\frac{1}{\gamma}\Big(\mathcal{E}_{\textrm{H-CP}}^{(m)}\Big)\tensor{}{_\alpha^\beta^\sigma} \dd \tensor{Q}{^\alpha_\beta_\sigma}+\frac{1}{\gamma} \d\Big(\dd\iota_{\vec{q}}\vol\Big)\\
    &\dd(\overline{\ell}^{(m)}_{\textrm{HMS}}-\overline{\ell}^{(m)}_{\textrm{PT}})-\jmath^*(\Theta^{(m)}_{\mathrm{HMS}}-\Theta^{(m)}_{\mathrm{PT}})=0
\end{align*}
giving non-trivial equations of motions. This shows that the HMS Lagrangians and Palatini Lagrangians are not in the same relative cohomology class.

\subsection{Definition of \texorpdfstring{$\xi$}{xi}-charges}\label{subsection xi-charges}
In this subsection we quicky summarize the definition of $\xi$-charges (we follow the notations, definitions, and results of \cite{CPS}). Consider a pair of Lagrangians $(L,\overline{\ell})$ over the space of fields $\mathcal{F}=\{\phi\ \text{tensor field}\}$ defining a good action principle. This means that we have
\[\dd L=E\wwedge\dd\phi+\d\Theta\qquad\qquad\dd\overline{\ell}-\jmath^*\Theta=\overline{b}\wwedge\dd\phi-\d\overline{\theta} \]
Given some vector field $\xi^\alpha$ tangent to $\partial_L M$, we define the $\xi$-currents and the $\xi$-charge associated with a Lagrangian pair $(L,\overline{\ell})$ as
\[
\begin{array}{l}
J_\xi:=\iota_\xi L-\ii_{\XX_\xi}\Theta\\
\overline{\jmath}_\xi:=-\iota_{\overline{\xi}}\overline{\ell}-\ii_{\XX_\xi}\overline{\theta}
\end{array}
\qquad\qquad\qquad\QQ^{\imath}_{\xi}:=\int_{(\Sigma,\partial\Sigma)}\underline{\imath}^*(J_\xi,\overline{\jmath}_\xi)\] 
where $\ii$ is the interior product of $\mathcal{F}$, $\dd\phi(\XX_\xi)=\L_\xi\phi$, and $\underline{\imath}:(\Sigma,\partial\Sigma)\hookrightarrow(M,\partial_LM)$ is a Cauchy embedding. The $\xi$-charges depend on the embedding off-shell but are independent on-shell.  Moreover, in general they also depend on the Lagrangians  chosen within the cohomological class. However, if we only allow Diff-invariant representatives, the charge does not depend on the choice.\vspace*{2ex}

Finally, lemma III.54 of \cite{CPS} shows that
\begin{equation}\label{eq: Hamilton eq}
	\dd\QQ^{\imath}_\xi=\ii_{\XX_\xi}\OOmega^\imath+\int_{(\Sigma,\partial\Sigma)}\underline{\imath}^*\Big(\underline{\imath}_\xi(E,\overline{b})\wedge\dd\phi\Big)+\int_{(\Sigma,\partial\Sigma)}\underline{\imath}^*(\underline{\L}_\xi-\underline{\LL}_{\widetilde{\XX}_\xi})\big(\Theta,\overline{\theta}\big)
	\end{equation}
If we have, as we do in our case, Diff-invariant representatives $(\Theta,\overline{\theta})$, the last integral is zero. Meanwhile, if we restrict ourselves to the space of solutions $\jj:\Sol\hookrightarrow\mathcal{F}$, then the first integral vanishes. Thus we have that $\XX_\xi|_{\Sol}$ is a Hamiltonian vector field with Hamiltonian $\QQ_\xi:=\jj^*\QQ^{\imath}_\xi$ ($\QQ_\xi$ is said to be integrable).

\subsection{HMS vs Palatini vs GR \texorpdfstring{$\xi$}{xi}-charges}

Let us prove that the  HMS and Palatini $\xi$-charges coincide and that they are both equal to the $\xi$-charges of GR after pulling back to the metric sector.\vspace*{2ex}

From \eqref{eq: Theta CP}, \eqref{theta_HMS_m} and \cite{CPSPT}, we have
\begin{align}
    \begin{split}
    (\Theta^{(m)}_{\mathrm{HMS}},\overline{\theta}^{(m)}_{\mathrm{HMS}})&=(\Theta^{(m)}_{\mathrm{PT}},\overline{\theta}^{(m)}_{\mathrm{PT}})+\frac{1}{\gamma}\underline{\dd}(\iota_{\vec{q}}\vol,0)\\
        &=(\Theta^{(m)}_{\mathrm{GR}},\overline{\theta}^{(m)}_{\mathrm{GR}})+\underline{\dd}(\iota_{\vec{A}-\Vec{C}+\vec{q}/\gamma}\vol,0)
    \end{split}
\end{align}
and
\begin{align}\begin{split}
    \big(L^{(m)}_{\textrm{HMS}},\overline{\ell}^{(m)}_{\textrm{HMS}}\big)&=(L^{(m)}_{\textrm{PT}},\overline{\ell}^{(m)}_{\textrm{PT}})-\frac{1}{\gamma}\Big(\hat{L}^{(m)}_{\textrm{H-CP}},0\Big)+\frac{1}{\gamma}\underline{\d}(\iota_{\vec{q}}\vol,0)\\
    &=\big(L^{(m)}_{\textrm{EH}},\overline{\ell}^{(m)}_{\textrm{GHY}}\big)+\Big(\hat{L}^{(m)}_{\textrm{P-CP}}-\frac{1}{\gamma}\hat{L}^{(m)}_{\textrm{H-CP}},0\Big)+\underline{\d}(\iota_{\vec{A} - \vec{C}+\vec{q}/\gamma}\vol,0)
\end{split}
\end{align} 
Then we have
\begin{align*}
&\QQ^{\mathrm{HMS},\imath}_{\xi,(m)}-\widetilde{\QQ}^{\mathrm{GR},\imath}_{\xi,(m)}=\int_{(\Sigma,\partial\Sigma)}\underline{\imath}^*\!\left(\underline{\iota}_\xi(L_{\mathrm{HMS}}^{(m)},\overline{\ell}_{\mathrm{HMS}}^{(m)})-\underline{\ii}_{\XX_\xi}(\Theta^{(m)}_{\mathrm{HMS}},\overline{\theta}^{(m)}_{\mathrm{HMS}})-\underline{\iota}_\xi(L_{\mathrm{EH}}^{(m)},\overline{\ell}_{\mathrm{GHY}}^{(m)})+\underline{\ii}_{\XX_\xi}(\Theta^{(m)}_{\mathrm{GR}},\overline{\theta}^{(m)}_{\mathrm{GR}})\right)\\
&=\int_\Sigma\imath^*\iota_\xi\Big(\hat{L}^{(m)}_{\textrm{P-CP}}-\frac{1}{\gamma}\hat{L}^{(m)}_{\textrm{H-CP}}\Big)+\int_{(\Sigma,\partial\Sigma)}\underline{\imath}^*\left(\underline{\iota}_\xi\underline{\d}(\iota_{\vec{A} - \vec{C}+\vec{q}/\gamma}\vol,0)-\underline{\ii}_{\XX_\xi}\underline{\dd}(\iota_{\vec{A} - \vec{C}+\vec{q}/\gamma}\vol,0)\right)\\
&=\int_\Sigma\imath^*\iota_\xi\Big(\hat{L}^{(m)}_{\textrm{P-CP}}-\frac{1}{\gamma}\hat{L}^{(m)}_{\textrm{H-CP}}\Big)+\int_{(\Sigma,\partial\Sigma)}\underline{\imath}^*(\underline{\L}_\xi-\underline{\d}\,\underline{\iota}_\xi)\underline{\iota}_{\vec{A} - \vec{C}+\vec{q}/\gamma}(\vol,0)-\int_{\Sigma}\imath^*(\LL_{\XX_\xi}-\dd\ii_{\XX_\xi})\iota_{\vec{A} - \vec{C}+\vec{q}/\gamma}\vol\\
&=\int_\Sigma\imath^*\iota_\xi\Big(\hat{L}^{(m)}_{\textrm{P-CP}}-\frac{1}{\gamma}\hat{L}^{(m)}_{\textrm{H-CP}}\Big)+\int_{\Sigma}\imath^*(\L_\xi-\LL_{\XX_\xi})\iota_{\vec{A} - \vec{C}+\vec{q}/\gamma}\vol=\int_\Sigma\imath^*\iota_\xi\Big(\hat{L}^{(m)}_{\textrm{P-CP}}-\frac{1}{\gamma}\hat{L}^{(m)}_{\textrm{H-CP}}\Big)
\end{align*}
To get the last line we have used the relative Stokes' theorem together with $\underline{\partial}(\Sigma,\partial\Sigma)=\varnothing$ and the fact that $\ii=0$ over $0$-forms over the space of fields. Finally, the last equality follows because $\iota_{\vec{A} - \vec{C}+\vec{q}/\gamma}\vol$ does not depend on any background object (over such objects $\LL_{\XX_\xi}=\L_\xi$). Finally, (see \eqref{eq: tilde=LC+}), we use that the coupling Lagrangians are zero on-shell to prove the equality of the charges.\vspace*{2ex}

Notice that we have written  $\widetilde{\QQ}_{\xi,(m)}^{\mathrm{GR},\imath}$, instead of simply  $\QQ_{\xi,(m)}^{\mathrm{GR},\imath}$, to remind the reader that, although they have the same functional expression, they live in different spaces. The former lives in the Palatini/HMS space $\mathcal{F}^{(m)}_{\textrm{PT}}(=\mathcal{F}^{(m)}_{\textrm{HMS}})$ while the latter lives in the GR space $\mathcal{F}^{(m)}_{\textrm{GR}}$ (they are equal after pullback/projection).\vspace*{2ex}

Since $\QQ_{\xi,(m)}^{\mathrm{HMS}}:=\jj^*\QQ_{\xi,(m)}^{\mathrm{HMS},\imath}=\jj^*\widetilde{\QQ}^{\mathrm{GR},\imath}_{\xi,(m)}$ does not depend on $\gamma$, we obtain $\QQ_{\xi,(m)}^{\mathrm{HMS}}=\QQ_{\xi,(m)}^{\mathrm{PT}}=\widetilde{\QQ}_{\xi,(m)}^{\mathrm{PT}}$ and all the charges are equivalent (the particular expression is given in \cite[III.4]{CPSGR}). Moreover, from \eqref{eq: Hamilton eq} it follows that these charges are the Hamiltonian of $\XX_\xi$
\begin{equation}\label{eq: Hamiltonian eq charges}
\ii_{\XX_\xi}\OOmega_{(m)}^{\mathrm{HMS}}=\dd\QQ^{\mathrm{HMS}}_{\xi,(m)}\,,\qquad\qquad\ii_{\XX_\xi}\OOmega^{\mathrm{PT}}_{(m)}=\dd\QQ^{\mathrm{PT}}_{\xi,(m)}\,,\qquad\qquad\ii_{\XX_\xi}\OOmega^{\mathrm{GR}}_{(m)}=\dd\QQ^{\mathrm{GR}}_{\xi,(m)}\,.
\end{equation}

\section{CHARGES IN THE TETRAD FORMALISM}
\subsection{HMS and Palatini Lagrangian pairs are NOT equal in relative cohomology}
We have seen that  $(L^{(m)}_{\textrm{HMS}},\overline{\ell}^{(m)}_{\textrm{HMS}})$ and $(L^{(m)}_{\textrm{PT}},\overline{\ell}^{(m)}_{\textrm{PT}})$ are different in the relative cohomology but, as mentioned on remark \ref{remark: dd Phi derivatives}, this does not imply that the tetrad counterparts $(L^{(t)}_{\textrm{HMS}},\overline{\ell}^{(t)}_{\textrm{HMS}})$ and $(L^{(t)}_{\textrm{PT}},\overline{\ell}^{(t)}_{\textrm{PT}})$ are different in relative cohomology. However, proceeding as in section \ref{subsection: pairs cohomology metric} allows us to show that they are different in the relative cohomology as well.

\subsection{HMS vs Palatini vs GR \texorpdfstring{$\xi$}{xi}-charges}

Let us prove that the  HMS and Palatini $\xi$-charges are equal and that they are both equal, after pulling back to the metric sector, to the $\xi$-charges of GR.\vspace*{2ex}

To this end first notice that
\begin{equation*}
    \widetilde{\omega}_{IJ} = \LC{\omega}_{IJ}+C_{IJ} + S_{IJ}\,,
\end{equation*}
The Lagrangian pairs of the three theories are related by
\begin{align}
    \begin{split}
    (L^{(t)}_{\textrm{HMS}},\overline{\ell}^{(t)}_{\textrm{HMS}})&=(L^{(t)}_{\textrm{PT}},\overline{\ell}^{(t)}_{\textrm{PT}})+\frac{1}{\gamma}\Big(e^I\wedge e^J\wedge(C_{IM}\wedge\tensor{C}{^M_J}+S_{IM}\wedge\tensor{S}{^M_J}),0\Big)+\\
    &\hspace*{4ex}+\frac{1}{\gamma}\underline{\d}\big(e^I\wedge e^J\wedge C_{IJ},0\big)=\\
    &=(L^{(t)}_{\textrm{EH}},\overline{\ell}^{(t)}_{\textrm{HGY}})+\frac{1}{2}\big(H_{IJKL}e^I\wedge e^J\wedge(\tensor{C}{^K_M}\wedge\tensor{C}{^M^L}+\tensor{S}{^K_M}\wedge\tensor{S}{^M^L}),0\big)+\\
    &\hspace*{4ex}+\frac{1}{2}\underline{\d}\big(H_{IJKL}e^I\wedge e^J\wedge C^{KL},0\big)\,,
    \end{split}
\end{align}
while their symplectic potentials are related by
\begin{align}\label{eq: Theta HMS PT GR}
    \begin{split}
    (\Theta^{(t)}_{\mathrm{HMS}},\overline{\theta}^{(t)}_{\mathrm{HMS}})&=(\Theta^{(t)}_{\mathrm{PT}},\overline{\theta}^{(t)}_{\mathrm{PT}})+\frac{1}{\gamma}(e^I\wedge e^J\wedge\dd C_{IJ},0)-\frac{1}{\gamma}\underline{\d}(e_I\wedge \dd e^I,0)\\
        &=(\Theta^{(t)}_{\mathrm{GR}},\overline{\theta}^{(t)}_{\mathrm{GR}})+\frac{1}{2}(H_{IJKL}e^I\wedge e^J\wedge\dd C^{KL},0)-\frac{1}{\gamma}\underline{\d}(e_I\wedge \dd e^I,0)\,.
    \end{split}
\end{align}
Using again the relative Stokes' theorem, the fact that there are no background objects and the relative Cartan's magic formula, we get

\begin{align*}
&\QQ^{\mathrm{HMS},\imath}_{\xi,(t)}-\widetilde{\QQ}^{\mathrm{GR},\imath}_{\xi,(t)}=\int_{(\Sigma,\partial\Sigma)}\underline{\imath}^*\!\left(\underline{\iota}_\xi(L_{\mathrm{HMS}}^{(t)},\overline{\ell}_{\mathrm{HMS}}^{(t)})-\underline{\ii}_{\XX_\xi}(\Theta^{(t)}_{\mathrm{HMS}},\overline{\theta}^{(t)}_{\mathrm{HMS}})-\underline{\iota}_\xi(L_{\mathrm{EH}}^{(t)},\overline{\ell}_{\mathrm{GHY}}^{(t)})+\underline{\ii}_{\XX_\xi}(\Theta^{(t)}_{\mathrm{GR}},\overline{\theta}^{(t)}_{\mathrm{GR}})\right)\\
&=\int_\Sigma\imath^*\left(\frac{1}{2}\iota_\xi\big(H_{IJKL}e^I\wedge e^J\wedge(\tensor{C}{^K_M}\wedge\tensor{C}{^M^L}+\tensor{S}{^K_M}\wedge\tensor{S}{^M^L})+H_{IJKL}(\L_\xi e^I)\wedge e^J\wedge C^{KL}\right)\,.
\end{align*}
On shell, $C_{IJ}=0$ and $\tensor{S}{^I_J}$ is proportional to the identity $\eta^I_J$. Thus, on shell, the previous expression vanishes and we have once again the equivalence of the charges (the particular expression is given in \cite[IV.8]{CPSGR}) and the fact that these charges are the Hamiltonian of $\XX_\xi$.\vspace*{2ex}

Once again, we have reached the expected equivalence between the metric and tetrad formalism. However, it is interesting to note that in the tetrad formalism we have an additional symmetry which is absent in the metric formalism. To introduce it, we need to take a small detour and speak about Kosmann derivatives, $\d$-symmetries, and general charges.

\subsection{Kosmann \texorpdfstring{$\xi$}{xi}-charges}
In the metric formalism, $g$-Killing vector fields $\xi\in\mathfrak{X}(M)$ satisfy $\L_\xi g=0$. If the metric is written in terms of tetrads as $g=\eta_{IJ}e^I\otimes e^J$, the condition $\L_\xi g=0$ is equivalent to requiring that $\L_{\xi} e^{I}=-\tensor{\hat{\lambda}}{^I_J}e^J$ for some $\tensor{\hat{\lambda}}{^{IJ}}$ antisymmetric but non-zero in general. This observation leads to the introduction of the Kosmann derivative  $\mathcal{K}_{\xi}e^{I}:=\L_{\xi} e^{I}+\tensor{\hat{\lambda}}{^I_J}e^J$, choosing $\hat{\lambda}^{IJ}$ so that it vanishes for every Killing vector field. The Kosmann derivative has been used, among other things, to study black hole entropy \cite{jacobson2015black} (see also \cite{Fatibene:1996tf} for a more geometric discussion).\vspace*{2ex}

In the bicomplex formalism, this construction can be defined as follows: consider a $\hat{\lambda}$-dependent vector field $\ZZ_{\hat{\lambda}}$ given by $\LL_{\ZZ_{\hat{\lambda}}}e^I:=\tensor{\hat{\lambda}}{^I_J}e^J$  and such that
\begin{equation}\label{eq: lambda xi e =0}
\LL_{\XX_\xi+\ZZ_{\hat{\lambda}}}e^I=0\qquad\equiv\qquad \L_\xi e^I+\tensor{\hat{\lambda}}{^I_J}e^J=0
\end{equation}
for every Killing vector field $\xi$. By demanding that \eqref{eq: lambda xi e =0} holds, we can get the explicit form of $\hat{\lambda}^{IJ}$ in the following way

\begin{align*}
\tensor{\hat{\lambda}}{^{IK}}&=\tensor{\hat{\lambda}}{^{[IK]}}=\eta^{J[K}\tensor{\hat{\lambda}}{^{I]}_J}=e_\alpha^JE^{\alpha [K}\tensor{\hat{\lambda}}{^{I]}_J}=-E^{\alpha[K}(\mathcal{L}_\xi e^{I]})_\alpha=-E^{\alpha[K}\left(\LC{\nabla}_{\vec{\xi}}\, e_\alpha^{I]}+e_\beta^{I]}\LC{\nabla}_\alpha \xi^\beta  \right)\\
&=\xi^\beta e_\alpha^{[I}\LC{\nabla}_\beta E^{\alpha K]}-E^{\alpha K}E^{\beta I}\LC{\nabla}_{[\alpha}\xi_{\beta]}=\imath_\xi \LC{\omega}^{IK}-\frac{1}{2}E^{\alpha K}E^{\beta I}(\mathrm{d}\xi)_{\alpha\beta}\,,
\end{align*}
where $\LC{\nabla}$ is the Levi-Civita connection. With the previous choice of $\hat{\lambda}^{IJ}$, we have:
\[(\LL_{\XX_\xi+\ZZ_{\hat{\lambda}}}e^I)_\alpha=\frac{1}{2}E^{I\beta}(\L_\xi g)_{\alpha\beta}\]
which indeed vanishes if $\xi$ is a Killing of the metric.\vspace*{2ex}

Now we have to prove that $\ZZ_{\hat{\lambda}}$ is a symmetry and compute the associated charges.

\subsection{Definition of \texorpdfstring{$\XX$}{XX}-charges}
A $\d$-symmetry is a vector field $\XX$ over $\mathcal{F}$ such that $\LL_\XX L$ is exact when no boundary is considered. If we assume that the base manifold has a non-empty boundary, then we have to rely on the concept of relative $\d$-symmetry (or $\underline{\d}$-symmetry for short) introduced in \cite{CPS}: a vector field $\XX$ over $\mathcal{F}$ such that $\underline{\LL}_\XX (L,\overline{\ell})$ is relative exact. That means that there exists some $(S_\XX,\overline{s}_\XX)$ such that
\[\underline{\LL}_\XX (L,\overline{\ell})=\underline{\d}(S_\XX,\overline{s}_\XX):=(\d S_\XX,\jmath^*S_\XX-\d\overline{s}_\XX)\,.\]
Now we can define the $\XX$-currents and $\XX$-charge
\[
(J_\XX,\overline{\jmath}_\XX):=(S_\XX,\overline{s}_\XX)-\underline{\ii}_{\XX}(\Theta,\overline{\theta})\,,\qquad\qquad\qquad\QQ^{\imath}_{\XX}:=\int_{(\Sigma,\partial\Sigma)}\underline{\imath}^*(J_\XX,\overline{\jmath}_\XX)\,.\] 
The charge $\QQ^\imath_\XX$ is independent of the chosen representatives. Moreover, restricting ourselves to the space of solutions we see that it does not depend on the embedding and it is the Hamiltonian of $\XX$. Finally, a comment is in order: if we apply these definitions to $\XX_\xi$, we recover the ones given in section \ref{subsection xi-charges} (although notice that those are defined even if $\XX_\xi$ is not a $\underline{\d}$-symmetry).\vspace*{2ex}

\subsection{Definition of \texorpdfstring{$\lambda$}{lambda}-charges}
Now we have to prove that $\ZZ_{\hat{\lambda}}$ is a $\underline{d}$-symmetry. In fact, we are going to prove something stronger, that $\ZZ_{\lambda}$ is a $\underline{d}$-symmetry for \textit{every} scalar field $\lambda^{IJ}$ antisymmetric in its internal indices. Let us define the vector field $\ZZ_\lambda$ over $\mathcal{F}^{(t)}_{\mathrm{PT}}$ by
\[\LL_{\ZZ_\lambda}e^I=\lambda^I_Je^J\,,\qquad\qquad\LL_{\ZZ_\lambda}\widehat{\omega}^{IJ}=-\widehat{D}\lambda^{IJ}\,,\qquad\qquad\LL_{\ZZ_\lambda}S^{IJ}=-\lambda^{IK}\tensor{S}{_K^J}-\lambda^{JK}\tensor{S}{_K^I}\]
From \eqref{eq: Phi lambda-inv} and \eqref{eq: L HMS}, it is clear that $\ZZ_\lambda$ is a $\underline{\d}$-symmetry of $(L^{(t)}_{\mathrm{HMS}},\overline{\ell}^{(t)}_{\mathrm{HMS}})$ and we can take $(S_{\ZZ_\lambda},\overline{s}_{\ZZ_\lambda})=(0,0)$. We now compute the $\ZZ_\lambda$-currents to obtain
\begin{align*}
    &J_{\ZZ_\lambda}=-\ii_{\ZZ_\lambda}\Theta^{(t)}_{\mathrm{HMS}}=\frac{1}{2}\d(H_{IJKL}\lambda^{KL}e^I\wedge e^J)-H_{IJKL}\lambda^{KL}\tensor{C}{^I_R}\wedge e^R\wedge e^J\,,\\
    &\overline{\jmath}_{\ZZ_\lambda}=-\ii_{\ZZ_\lambda}\overline{\theta}^{(t)}_{\mathrm{HMS}}=\varepsilon_{IJKL}\lambda^{RJ}\overline{e}^K\wedge\overline{e}^L N^I N_R+\frac{1}{\gamma}\lambda^{IJ}\overline{e}_I\wedge \overline{e}_J\,.
\end{align*}
These expressions can be rewritten in a relative form as
\begin{align*}
    (J_{\ZZ_\lambda},\overline{\jmath}_{\ZZ_\lambda})&=\frac{1}{2}\underline{\d}\big(H_{IJKL}\lambda^{KL}e^I\wedge e^J,0\big)-\big(H_{IJKL}\lambda^{KL}\tensor{C}{^I_R}\wedge e^R\wedge e^J,0\big){}+{}\\
    &\phantom{=}+\frac{1}{2}\big(0,\varepsilon_{IJKL}(2 N^I N_R\lambda^{RJ}-\lambda^{IJ})\overline{e}^K\wedge\overline{e}^L\big)\,.
\end{align*}
Finally, notice that the last term is zero as a consequence of $\epsilon_{[IJKL}N_{R]}N^I\lambda^{JR}\overline{e}^K\wedge\overline{e}^L=0$.\vspace*{2ex}

We end by computing the $\ZZ_\lambda$-charge ($\lambda$-charge for short) as the integral of the $\ZZ_\lambda$-currents
\[\QQ^{\mathrm{HMS},\imath}_\lambda:=\int_{(\Sigma,\partial\Sigma)} (J_{\ZZ_\lambda},\overline{\jmath}_{\ZZ_\lambda})=-\int_\Sigma H_{IJKL}\lambda^{KL}\tensor{C}{^I_R}\wedge e^R\wedge e^J\,,\]
which is zero  on-shell (since  over the space of solutions $C^{IJ}:=\widehat{\omega}^{IJ}-\LC{\omega}^{IJ}$ vanishes). Moreover, from \eqref{eq: Theta HMS PT GR} and the relative Stokes' theorem, we have
\[\QQ^{\mathrm{HMS},\imath}_\lambda=\QQ^{\mathrm{PT},\imath}_\lambda-\frac{1}{\gamma}\int_\Sigma e^I\wedge e^J\wedge\LL_{\ZZ_\lambda}C_{IJ}=
\QQ^{\mathrm{GR},\imath}_\lambda-\frac{1}{2}\int_\Sigma H_{IJKL}e^I\wedge e^J\wedge\LL_{\ZZ_\lambda} C^{KL}\,.\]
We see that the $\lambda$-charges do not coincide off-shell in GR, Palatini, and HMS. However, as expected since there is no metric counterpart, they all vanish on-shell.

\section{CONCLUSIONS AND COMMENTS}\label{sec:CONCLUSIONS}

In this paper we have studied in full detail the relation between the metric-HMS and tetrad-HMS formulations for general relativity on manifolds with or without boundary. First we have proven that the spaces of solutions of the metric-HMS action and the metric-Palatini action are the same. Then we have studied the correspondence between the solution spaces in the metric and tetrad formalisms. Although the simple relationship between them can be justified on general grounds by relying on the properties of the transformation $\Phi_{\mathrm{HMS}}$ and its tangent map (in particular, by the fact that both are onto), we have checked this explicitly by solving the relevant sector of the field equations. In order to do this, we have used the irreducible decompositions of the tensors involved. We would like to insist on several facts:

\begin{itemize}
\item We have done this in full generality, i.e. by taking from the start complete general connections with torsion and non metricity. In particular, in the tetrad formalism we have the Holst action plus another term that depends on the symmtric part of the connection. To the best of our knoweledge, this has not been considered before.
\item We have derived a new boundary Lagrangian to recover GR also at the boundary, and used the transformation $\Phi_{\mathrm{HMS}}$ to find its tetrad counterpart. The latter coincides with the boundary Lagrangian proposed in \cite{bodendorfer2013imaginary} by Bodendorfer and Neiman (although they only work with Lorentz connections).
\item As expected, the equivalence of the solution spaces extends to the case of manifolds with boundaries.
\end{itemize}

As far as the (pre)symplectic forms are concerned the situation is very simple in the metric case as the symplectic potential corresponding to the different formulations (Einstein-Hilbert, Palatini, and HMS) differ by a $\dd$-exact term. In fact, they coincide both off-shell and on-shell. The tetrad case is more complicated. This is to be expected on general grounds because the transformation $\Phi_{\mathrm{HMS}}$ involves derivatives. As we have shown, the HMS and Palatini symplectic potentials are not equal on the relative cohomology: they are different off-shell but coincide on-shell as a consequence of the dynamical vanishing of the contortion $C_{IJ}$.\vspace*{2ex}

Finally, regarding the charges we have shown that they also differ off-shell, but coincide on shell (again as a consequence of the fact that $C_{IJ}=0$). A similar analysis has been performed for the $\lambda$-charges (which include the Kosmann charges) proving that, in fact, they all vanish.

\begin{acknowledgments}
This work has been supported by the Spanish Ministerio de Ciencia Innovaci\'on y Uni\-ver\-si\-da\-des-Agencia Estatal de Investigaci\'on FIS2017-84440-C2-2-P and PID2020-116567GB-C22 grants. Juan Margalef-Bentabol is supported by the AARMS postdoctoral fellowship, by the NSERC Discovery Grant No. 2018-04873, and the NSERC Grant RGPIN-2018-04887. E.J.S. Villase\~nor is supported by the Madrid Government (Comunidad de Madrid-Spain) under the Multiannual Agreement with UC3M in the line of Excellence of University Professors (EPUC3M23), and in the context of the V PRICIT (Regional Programme of Research and Technological Innovation). We have made extensive use of the xAct packages \cite{xact}. We thank José María Martín-García for his help.
\end{acknowledgments}

\newpage

\bibliographystyle{plainnat}
\bibliography{PRL}
\end{document}